%% file: TOP-24-001_temp.tex
\pdfoutput=1
\documentclass[11pt,twoside,a4paper,cmspaper,final,collab]{cms-tdr}
\def\svnVersion{5ed54d4}\def\svnDate{2026/05/20}\def\cmsCernNoTag{CERN-EP-2025-201}\def\cmsCernDate{\today}\def\cmsMessage{Published in the Journal of High Energy Physics as \href{http://dx.doi.org/10.1007/JHEP05(2026)175}{\doi{10.1007/JHEP05(2026)175}.}}
\begin{document}\cmsNoteHeader{TOP-24-001}

\providecommand{\cmsTable}[1]{\resizebox{\textwidth}{!}{#1}}
\newlength\cmsTabSkip\setlength{\cmsTabSkip}{1ex}

\newcommand{\mll}{\ensuremath{m_{\Pell\Pell}}\xspace}
\newcommand{\ttll}{\ensuremath{\Pell\Pell}\xspace}
\newcommand{\ee}{\EE}
\newcommand{\mumu}{\MM}
\newcommand{\emu}{\ensuremath{\Pepm\PGmmp}\xspace}

\newcommand{\pTnunu}{\ensuremath{\pt^{\nu\nu}}\xspace}
\newcommand{\pTVecnunu}{\ensuremath{\ptvec^{\kern1pt\nu\nu}}\xspace}
\newcommand{\pTVecl}{\ensuremath{\ptvec^{\kern1pt\Pell}}\xspace}
\newcommand{\ptmissdnn}{\ensuremath{p_{\mathrm{T,DNN}}^\text{miss}}\xspace}
\newcommand{\ptmisspuppi}{\ensuremath{p_{\mathrm{T,PUPPI}}^\text{miss}}\xspace}
\newcommand{\ptmisspf}{\ensuremath{p_{\mathrm{T,PF}}^\text{miss}}\xspace}
\newcommand{\ptmissgen}{\ensuremath{p_{\mathrm{T,gen.}}^\text{miss}}\xspace}
\newcommand{\ptmissrec}{\ensuremath{p_{\mathrm{T,rec.}}^\text{miss}}\xspace}
\newcommand{\ptvecmissdnn}{\ensuremath{\vec{p}_{\mathrm{T,DNN}}^{\kern1pt\text{miss}}}\xspace}
\newcommand{\ptvecmisspuppi}{\ensuremath{\vec{p}_{\mathrm{T,PUPPI}}^{\kern1pt\text{miss}}}\xspace}
\newcommand{\ptvecmisspf}{\ensuremath{\vec{p}_{\mathrm{T,PF}}^{\kern1pt\text{miss}}}\xspace}
\newcommand{\ptvecmissgen}{\ensuremath{\vec{p}_{\mathrm{T,gen.}}^{\kern1pt\text{miss}}}\xspace}
\newcommand{\ptvecmissrec}{\ensuremath{\vec{p}_{\mathrm{T,rec.}}^{\kern1pt\text{miss}}}\xspace}
\newcommand{\phiNextLepNuNuVec}{\ensuremath{\min[\Delta\phi(\pTVecnunu,\pTVecl)]}\xspace}
\newcommand{\phiNextLepMETVec}{\ensuremath{\min[\Delta\phi(\ptvecmissdnn,\pTVecl)]}\xspace}

\newcommand{\pp}{\ensuremath{\Pp\Pp}\xspace}
\newcommand{\WW}{\ensuremath{\PW\PW}\xspace}
\newcommand{\WZ}{\ensuremath{\PW\PZ}\xspace}
\newcommand{\ZZ}{\ensuremath{\PZ\PZ}\xspace}
\newcommand{\tW}{\ensuremath{\PQt\PW}\xspace}
\newcommand{\ttWZ}{\ensuremath{\ttbar\PW\!/\PZ}\xspace}
\newcommand{\chisq}{\ensuremath{\chi^2}\xspace}
\newcommand{\ndf}{\ensuremath{\text{ndf}}\xspace}
\newcommand{\chisqndf}{\ensuremath{\chisq\!/\ndf}\xspace}
\newcommand{\abseta}{\ensuremath{\abs{\eta}}\xspace}

\cmsNoteHeader{TOP-24-001}

\title{Measurement of the dineutrino system kinematic variables in dileptonic top quark pair production in proton-proton collisions at \texorpdfstring{$\sqrt{s}=13\TeV$}{sqrt(s) = 13 TeV}}

\date{\today}

\abstract{Differential top quark pair production cross sections are measured in the dilepton final states \ee, \mumu, and \emu, as a function of kinematic variables of the two-neutrino system: the transverse momentum \pTnunu of the dineutrino system, the minimum distance in azimuthal angle between \pTVecnunu and leptons, and in two dimensions in bins of both observables. The measurements are performed using CERN LHC proton-proton collisions at \texorpdfstring{$\sqrt{s}=13\TeV$}{sqrt(s) = 13 TeV}, recorded by the CMS detector between 2016 and 2018, corresponding to an integrated luminosity of 138\fbinv. The measured cross sections are unfolded to the particle level using an unregularized least squares method. Results are compared with predictions by the standard model of particle physics, and found to be in agreement with theoretical calculations as well as Monte Carlo simulations.}

\hypersetup{
pdfauthor={CMS Collaboration},
pdftitle={Measurement of the dineutrino system kinematic variables in dileptonic top quark pair production in proton-proton collisions at sqrt(s) = 13 TeV},
pdfsubject={CMS},
pdfkeywords={CMS, top quark, missing transverse energy, neutrinos, differential, cross sections}}

\maketitle

\section{Introduction}

Precision measurements of the top quark pair (\ttbar) production cross section provide stringent tests for the validity of the standard model of particle physics (SM) while playing a crucial role in the search for new phenomena. The large \ttbar data samples collected at the CERN LHC have made several measurements of differential cross sections possible in various \ttbar decay channels and at different center-of-mass energies~\cite{ATLAS:2012vvt,CMS:2012hkm,ATLAS:2014ipf,ATLAS:2015dbj,ATLAS:2016pal,CMS:2015rld,ATLAS:2015lsn,ATLAS:2015mip,CMS:2015auz,CMS:2016poo,CMS:2017iqf,ATLAS:2019guf,CMS:2016oae,ATLAS:2016qjg,ATLAS:2016pbv,CMS:2017xio,CMS:2018htd,CMS:2018adi,CMS:2019esx,ATLAS:2019hxz,ATLAS:2019hau,ATLAS:2020ccu,CMS:2021vhb}. All measurements to date were performed as functions of observables that characterize the kinematic properties of either the visible part of the event, such as jets or charged leptons, or intermediate particles, such as the top (anti)quark or the \PW boson. However, there are beyond-the-SM (BSM) physics scenarios that mostly modify the invisible part of the signature. Precise measurements of undetected particles in the event, can therefore reveal signs of new phenomena.

This paper presents the first measurements of the differential \ttbar cross section as a function of the transverse momentum of the dineutrino system, \pTnunu, and the minimum azimuthal distance between \pTVecnunu and the transverse momentum of a charged lepton, \phiNextLepNuNuVec, which can be precisely reconstructed in dileptonic final states. In addition, a two-dimensional (2D) measurement as a function of both observables is performed. For the SM \ttbar production process, neutrinos from the leptonic decays of the W boson account for most of the missing transverse momentum \ptvecmiss of the event, while for BSM processes, contributions to \ptvecmiss from undetected particles are also present in the event. These additional contributions can weaken the angular correlation between \pTVecnunu and the lepton direction, leading to a broader \phiNextLepNuNuVec distribution. The selection of the two observables for the cross section measurements in this analysis is, therefore, targeted to validate the modelling of the SM \ttbar process in a phase space sensitive to BSM scenarios that exhibit a comparable signature but include additional sources of undetected particles.

\begin{figure}[!htp]
\centering
\includegraphics[width=0.4\textwidth]{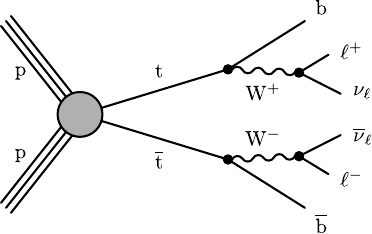}
\hspace*{0.05\textwidth}
\includegraphics[width=0.4\textwidth]{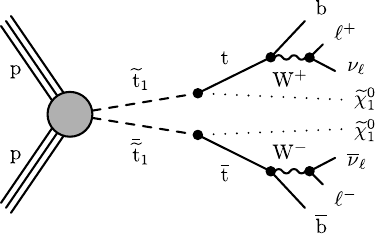}
\caption{In the left diagram, the SM \ttbar production process is sketched, whereas the right diagram shows the production of a hypothetical top squark pair, $\PSQtDo\PASQtDo$, with both top squarks decaying to a top quark and a neutralino, \PSGczDo. This analysis focuses on signatures, where both of the \PW bosons decay leptonically.}
\label{fig:diagrams}
\end{figure}

Squarks are superpartners of quarks as predicted by supersymmetry \cite{Weinberg:1981wj}. In Fig.~\ref{fig:diagrams}, a diagram of \ttbar production in the dilepton final state is compared to that of a top squark pair production, as an example of a BSM process \cite{Beenakker:2016gmf}. The presence of neutralinos, that neither decay to SM particles nor are detectable within the detector, modifies the kinematic properties of the invisible part of the BSM event that otherwise exhibits a final state identical to that of the \ttbar decay. Dedicated searches for top squark pair production in the dilepton final state found significant contributions of \ttbar events in the search phase space~\cite{CMS:2020pyk,CMS:2021eha,ATLAS:stopsearch}.

A precise measurement of the dineutrino system in dileptonic \ttbar events requires a good resolution of \ptvecmiss, which serves as a proxy for the transverse momentum of the dineutrino system. The magnitude and direction of \ptvecmiss are evaluated from the negative \ptvec sum of all reconstructed objects in an event, exploiting momentum conservation. The \ptvecmiss resolution is affected by mismeasurements of other particles that can happen due to detector effects and/or as a result of additional proton-proton (\pp) collisions in the same or adjacent bunch crossing (pileup). A dedicated correction based on a regression using machine learning methods is developed to improve the \ptvecmiss resolution.

Absolute and normalized differential cross sections are measured in the dilepton final states \ee, \mumu, and \emu,  derived for a fiducial phase space close to the acceptance of the detector and extracted by an unregularized unfolding using a least squares minimization.
The results are compared to predictions from Monte Carlo (MC) event generators at the matrix element (ME) level with next-to-leading order (NLO) accuracy in quantum chromodynamics (QCD), which are interfaced with a parton shower (PS) simulation. In addition, the results are compared to fixed-order predictions at NLO and next-to-NLO (NNLO) precision~\cite{Czakon:2020qbd}.

The paper is structured as follows: Section~\ref{sec:cms} briefly describes the main features of the CMS detector. In Section~\ref{sec:samples}, the recorded data samples are briefly introduced and settings for MC simulation of signal and background processes are detailed.
Section~\ref{sec:selection} presents the object reconstruction and the event selection, followed by the description of the correction of \ptvecmiss in Section~\ref{sec:dnn}. The systematic uncertainties and the results, including comparisons to different predictions, are discussed in Sections~\ref{sec:sysunc}~and~\ref{sec:results}, respectively. Finally, the analysis is summarised in Section~\ref{sec:summary}. Tabulated results can be found in a HEPData record~\cite{hepdata}.

\section{The CMS detector}
\label{sec:cms}

The CMS apparatus~\cite{CMS:2008xjf,CMS:2023gfb} is a multipurpose, nearly hermetic detector, designed to trigger on~\cite{CMS:2020cmk,CMS:2016ngn,CMS:2024aqx} and identify electrons, muons, photons, charged and neutral hadrons~\cite{CMS:2020uim,CMS:2018rym,CMS:2014pgm}. Its central feature is a superconducting solenoid of 6\unit{m} internal diameter, providing a magnetic field of 3.8\unit{T}. Within the solenoid volume are a silicon pixel and strip tracker, a lead tungstate crystal electromagnetic calorimeter (ECAL), and a brass and scintillator hadron calorimeter (HCAL), each composed of a barrel and two endcap sections. Forward calorimeters extend the pseudorapidity ($\eta$) coverage provided by the barrel and endcap detectors. Muons are reconstructed using gas-ionization detectors embedded in the steel flux-return yoke outside the solenoid. Events of interest are selected using a two-tiered trigger system. The first level, composed of custom hardware processors, uses information from the calorimeters and muon detectors to select events at a rate of around 100\unit{kHz} within a fixed latency of 4\mus~\cite{CMS:2020cmk}. The second level, known as the high-level trigger, consists of a farm of processors running a version of the full event reconstruction software optimized for fast processing, and reduces the event rate to around 1\unit{kHz} before data storage~\cite{CMS:2016ngn}.

More detailed descriptions of the CMS detector, together with a definition of the coordinate system used and the relevant kinematic variables, can be found in Refs.~\cite{CMS:2008xjf,CMS:2023gfb}.

\section{Data and simulated samples}
\label{sec:samples}

The measurements presented in this paper are based on \pp collision data recorded between 2016 and 2018 at a center-of-mass energy of 13 \TeV, corresponding to an integrated luminosity of 138\fbinv~\cite{CMS:2021xjt,CMS:2018elu,CMS:2019jhq} with 36.3, 41.5, and 59.8\fbinv of data recorded in 2016, 2017, and 2018, respectively.

Both the signal and the SM background processes are simulated using MC event generators. Simulated samples are produced separately for each data-taking period to match the corresponding pileup and detector conditions. All simulated samples use parton distribution functions (PDFs) as implemented in NNPDF3.1~\cite{Ball:2011uy,NNPDF:2017mvq} at NNLO accuracy in perturbative QCD. All ME event generators are interfaced with \PYTHIA v8.230~\cite{Sjostrand:2014zea} for the simulation of the parton showering and hadronization using the CP5 tune~\cite{CMS:2019csb} unless stated otherwise in the text. The detector response is simulated with \GEANTfour~\cite{GEANT4:2002zbu}.

The \ttbar production process is simulated using the \POWHEG v2~\cite{Nason:2004rx,Frixione:2007vw,Alioli:2010xd} event generator at NLO precision in QCD with the top quark mass set to 172.5\GeV. Top quark pairs in which both W bosons decay into either an electron or muon belong to the signal sample and are referred to as ``\ttbar (\ttll)'' and/or ``dileptonic \ttbar'' in the following. All other decay channels, including the production of one or more tau leptons, are denoted as ``\ttbar other''. Additional \ttbar samples using the same event generator are produced with modified parameters for the underlying event tune, color reconnection scheme, jet matching scheme, and top quark mass, and are used in the estimation of the impact of theoretical uncertainties. A \ttbar sample generated with \MGvATNLO v2.4.2 (MG5)~\cite{Alwall:2014hca} at NLO accuracy and interfaced with \PYTHIA through the FxFx~\cite{Frederix:2012ps} matching and merging scheme is used for the determination of corrections that improve the experimental \ptvecmiss resolution detailed in Section~\ref{sec:dnn}. This sample is referred to as \MCATNLO{}+\PYTHIA in the following. Another alternative \ttbar sample is generated with the \POWHEG v2 generator interfaced with \HERWIG v7.1~\cite{Bellm:2015jjp} using the CH3 tune~\cite{CMS:2020dqt}, and is referred to as \POWHEG{}+\HERWIG.

Background processes are modelled using different MC event generators. Drell--Yan events with additional jets and two leptons in the final state (DY+jets) are produced at NLO using MG5. Events are passed through PS and hadronization using the FxFx merging scheme.

The same event generator is used for the production of a \PW boson that subsequently decays to a lepton and neutrino. Events contain up to four additional jets, modelled at leading order in perturbative QCD, and matched to jets produced during PS and hadronization with the MLM scheme~\cite{Alwall:2007fs}.
The electroweak top quark productions, referred to as ``single top'' hereafter, are modelled at NLO accuracy using \POWHEG v2~\cite{Alioli:2009je} for the \tW process and the $t$-channel production, and using MG5 in the case of the $s$-channel production. For the \tW process, the description based on the diagram subtraction (DS) scheme is applied~\cite{Re:2010bp} in order to handle the overlap and interference with \ttbar production. Another \tW sample based on the diagram removal (DR) scheme~\cite{Re:2010bp} is used for the estimation of theoretical uncertainties.

The production of \ttbar in association with a \PW or \PZ boson (\ttWZ) is modeled at NLO accuracy using MG5. Diboson production processes, including \WW, \WZ, and \ZZ, are generated at leading order using \PYTHIA. The contribution of QCD multijet production in the analysis phase-space is negligible.

{\tolerance=800
The expected yield of \ttbar events is determined using a cross section calculated with the \textsc{top++}~\cite{Czakon:2011xx} program at NNLO plus next-to-next-to-leading-logarithmic accuracy~\cite{Czakon:2013goa,Cacciari:2011hy,Barnreuther:2012wtj,Czakon:2012zr}, assuming a top quark mass of 172.5\GeV. The simulated \pt spectra of top quarks are reweighted to NNLO cross sections using NNLO/NLO K-factors~\cite{Czakon:2017wor}. Yields from the {\PW}+jets and high-mass DY+jets processes are calculated using NNLO accuracy cross sections~\cite{Gavin:2010az,Li:2012wna}, while the yield of low-mass DY+jets and the \ttWZ processes are calculated using NLO accuracy cross sections from the MG5 generator. A cross section calculation at approximate NNLO is used for the single top \tW~\cite{Kidonakis:2010ux} process, while yields from the single top $s$- and $t$-channel processes are determined using NLO cross sections~\cite{Aliev:2010zk,Kant:2014oha}.
Cross sections calculated at NNLO accuracy~\cite{Gehrmann:2014fva} are used to determine the \WW yield, whereas NLO cross sections~\cite{Campbell:2011bn} are applied to determine the \WZ and \ZZ diboson process yields.
\par}

\section{Object reconstruction and event selection}
\label{sec:selection}

Events passing dilepton and single-lepton triggers are used for the offline analysis. The trigger criteria vary by data-taking period. For dielectron and electron-muon triggers, the transverse momentum \pt of the (sub)leading lepton is required to pass a threshold of 23 (12)\GeV. The maximum \pt threshold on the (sub)leading muon in dimuon triggers is 17 (8)\GeV. The threshold on the lepton \pt is 27 (32)\GeV in 2016 (2017--2018) for single electron triggers, and 24 (27)\GeV in 2016/2018 (2017) for single muon triggers.

The primary vertex (PV) is taken to be the vertex corresponding to the hardest scattering in the event, evaluated using tracking information alone, as described in Section 9.4.1 of Ref.~\cite{Contardo:2020886}.
The particle-flow (PF) algorithm~\cite{CMS:2017yfk} aims to reconstruct and identify each individual particle in the event, with an optimized combination of information from various elements of the CMS detector. The energy of photons is obtained from the ECAL measurement. The energy of electrons is determined from a combination of the electron momentum at the PV, the energy of the corresponding ECAL cluster, and the energy sum of all bremsstrahlung photons spatially compatible with originating from the electron track. The energy of muons is obtained from the curvature of the corresponding track. The energy of charged hadrons is determined from a combination of their momenta measured in the tracker and the matching ECAL and HCAL energy deposits, corrected for the response function of the calorimeters to hadronic showers. Finally, the energy of neutral hadrons is obtained from the corresponding corrected ECAL and HCAL energies.

Electron candidates are vetoed if their associated ECAL cluster is reconstructed within the transition region of the ECAL, which corresponds to the $\eta$ range of $1.44<\abseta<1.57$. Misidentified and nonprompt electrons are further suppressed by implementing selections based on discriminating quantities, such as shower shape variables, relative isolation, longitudinal and transverse impact parameters with respect to the PV, and photon conversion veto observables. For electrons, the relative isolation is defined as the \pt sum of all neutral and charged hadrons, and photon candidates within an angular distance $\DR=\sqrt{\smash[b]{(\Delta\phi)^2+(\Delta\eta)^2}}$ of 0.3 from the electron in the $\eta$--$\phi$ space, divided by the \pt of the electron candidate. Depending on the \pt and $\eta$ of the electron candidate, a maximum relative isolation in the range 0.05--0.1 is enforced. Muon candidates are required to have matching tracks in the inner tracker and the muon system, with track quality criteria, relative isolation, and impact parameter requirements suppressing nonprompt or misidentified muons. For muons, the relative isolation is based on contributions within a distance of 0.4 from the muon candidate, applying a maximal relative isolation of 0.15. For both electrons and muons, the relative isolation is corrected for pileup effects by removing charged PF candidates associated with pileup vertices, and by estimating the contribution of neutral particles from pileup to be subtracted from the \pt sum~\cite{CMS:2024ppo,CMS:2024qgs}.

Jets are clustered using the anti-\kt algorithm~\cite{Cacciari:2008gp, Cacciari:2011ma} with a distance parameter of 0.4. Jets within an angular distance $\DR<0.4$ of selected leptons are not considered in further steps of the analysis. To reduce contributions from jets originating from pileup vertices, additional requirements for jets reconstructed with $\pt<50\GeV$ are applied~\cite{CMS:2020ebo}. Jets originating from the hadronization of \PQb quarks are identified with the \textsc{DeepJet} algorithm~\cite{Bols:2020bkb}, using a working point corresponding to an efficiency of up to 92\% for \PQb quark jets and a misidentification rate of 10 (40--50)\% for jets originating from light-flavor quarks and gluons (\PQc quarks).

The missing transverse momentum vector \ptvecmiss is computed as the negative vector sum of the transverse momenta of all the PF candidates in an event, and its magnitude is denoted as \ptmiss~\cite{CMS:2019ctu}. The \ptvecmiss is modified to account for corrections to the energy scale of the reconstructed jets in the event and denoted as \ptvecmisspf. The pileup-per-particle identification (PUPPI) algorithm~\cite{Bertolini:2014bba} is applied to reduce the pileup dependence of the \ptvecmiss observable. For each PF candidate, a weight is computed based on the probability to originate from the PV. For charged particles, the assignment of the weight is based on the track-vertex association. For neutral PF candidates, instead, the weight computation is based on the \pt of the surrounding particles, as described in Ref.~\cite{CMS:2019ctu}. The \ptvecmiss computed from the PF candidates weighted by their probability to originate from the PV is denoted as \ptvecmisspuppi. Further corrections to \ptvecmisspuppi are derived in this analysis using a deep neural network (DNN) in order to improve the \ptmiss resolution in dileptonic \ttbar events, especially at high values of \ptmiss. This procedure is further detailed in Section~\ref{sec:dnn} and the corrected \ptvecmiss is denoted as \ptvecmissdnn. Anomalous high-\ptmiss events can occur due to a variety of failures in the object reconstruction, detector malfunctions, and/or noncollision backgrounds. Such events are rejected by event filters designed to identify more than 85--90\% of the spurious high-\ptmiss events with a misidentification rate of less than 0.1\%~\cite{CMS:2019ctu}.

The event selection closely follows that of Ref.~\cite{CMS:2018adi} with the exception that the requirement on missing transverse momentum is applied to \ptvecmissdnn rather than \ptvecmisspf. Events with two oppositely charged leptons that fulfill $\abseta<2.4$ and $\pt>25$ (20)\GeV for the leading (subleading) lepton are selected and split into the same-flavor {\ee}/\mumu and the different-flavor \emu channels. Events containing additional leptons (electrons or muons) with $\pt>15\GeV$ are rejected. Contributions from low-mass resonances decaying to two leptons are suppressed by requiring a minimal invariant mass \mll of the two leptons of 20\GeV. In order to reduce the contribution from DY+jet events, the events in the same-flavor channels are required to fulfill $\ptmissdnn>40\GeV$ and are vetoed if \mll is compatible with the \PZ boson mass~\cite{ParticleDataGroup:2024cfk} within 15\GeV. Selected events must contain at least two reconstructed jets with $\pt>30\GeV$ and $\abseta<2.4$, of which at least one is \PQb tagged ~\cite{CMS:2017wtu}.

\begin{figure}[!htb]
\centering
\includegraphics[width=0.48\textwidth]{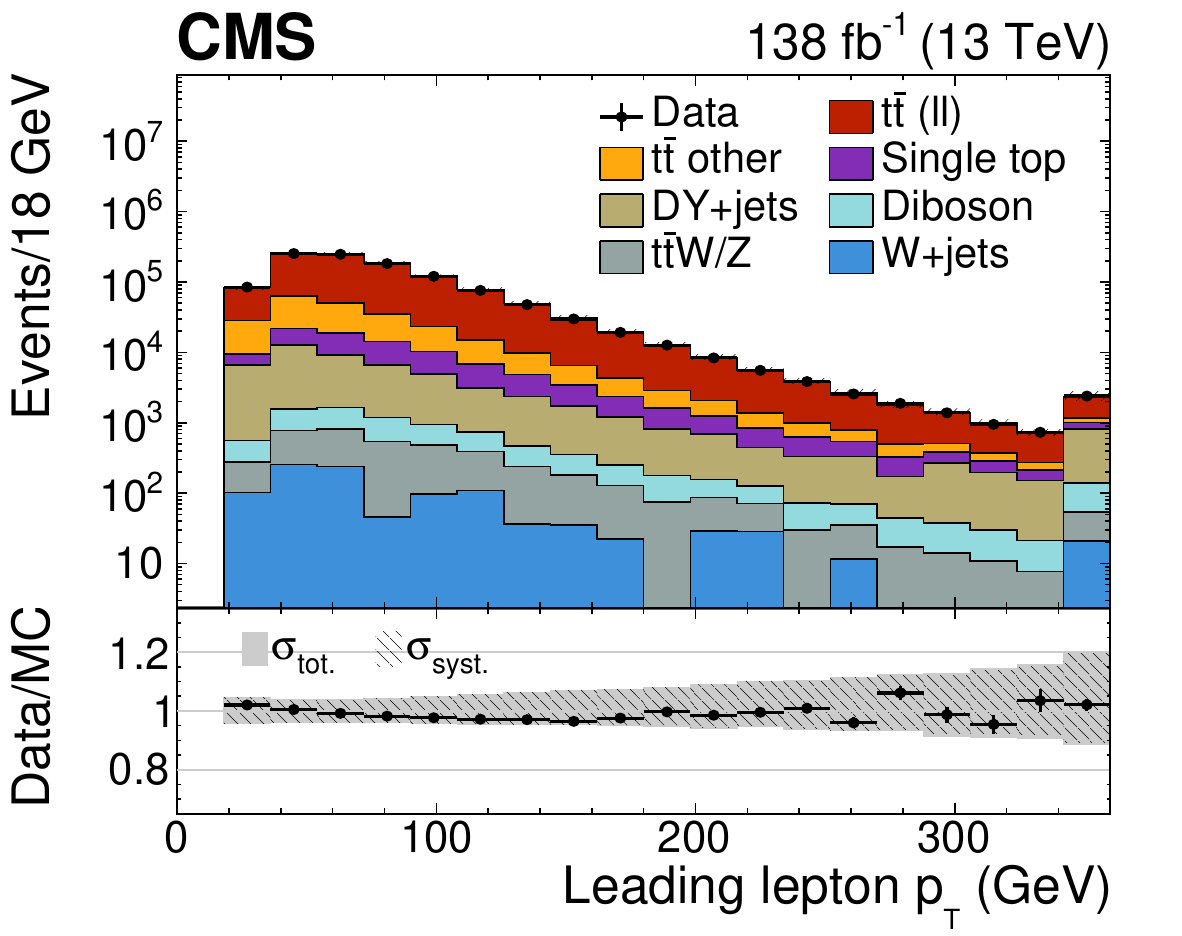}%
\hfill%
\includegraphics[width=0.48\textwidth]{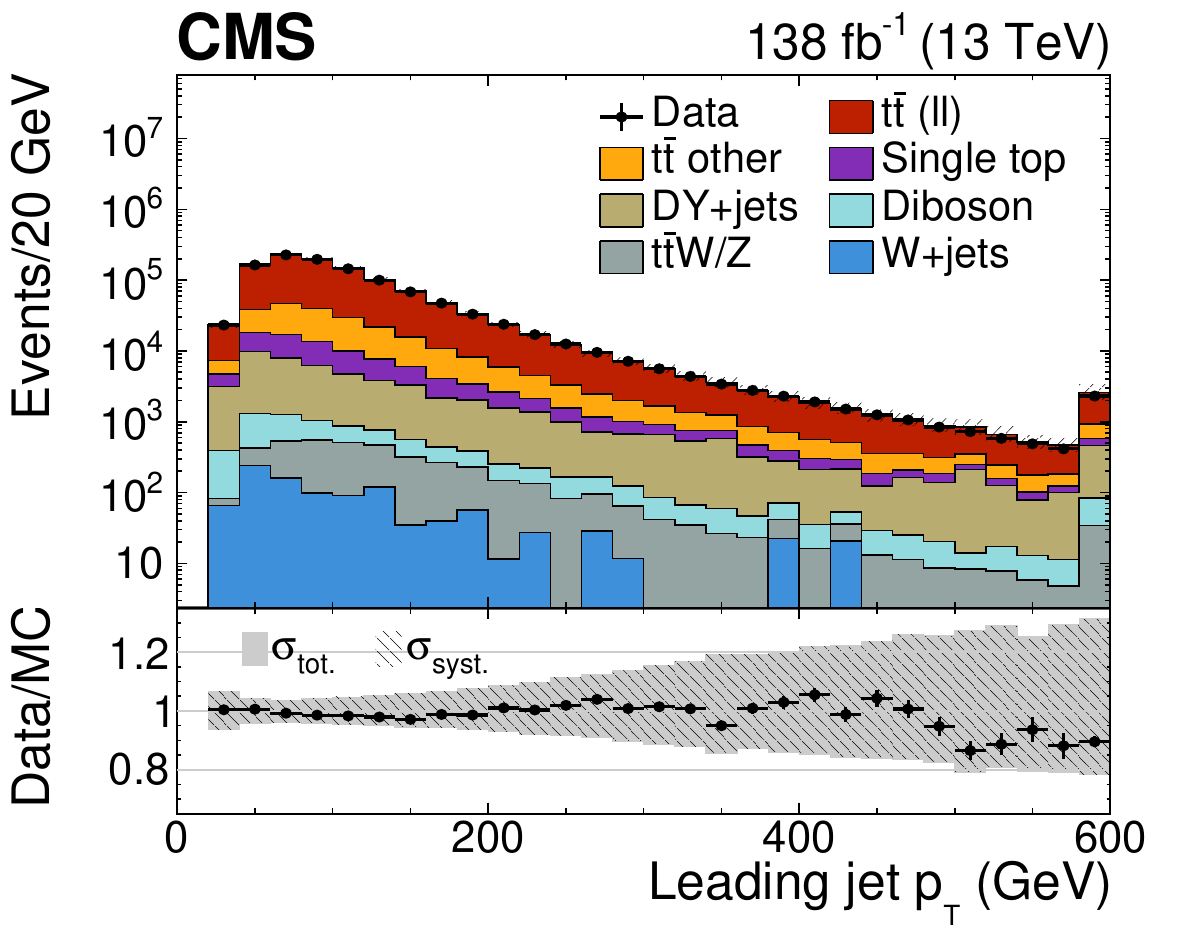} \\
\includegraphics[width=0.48\textwidth]{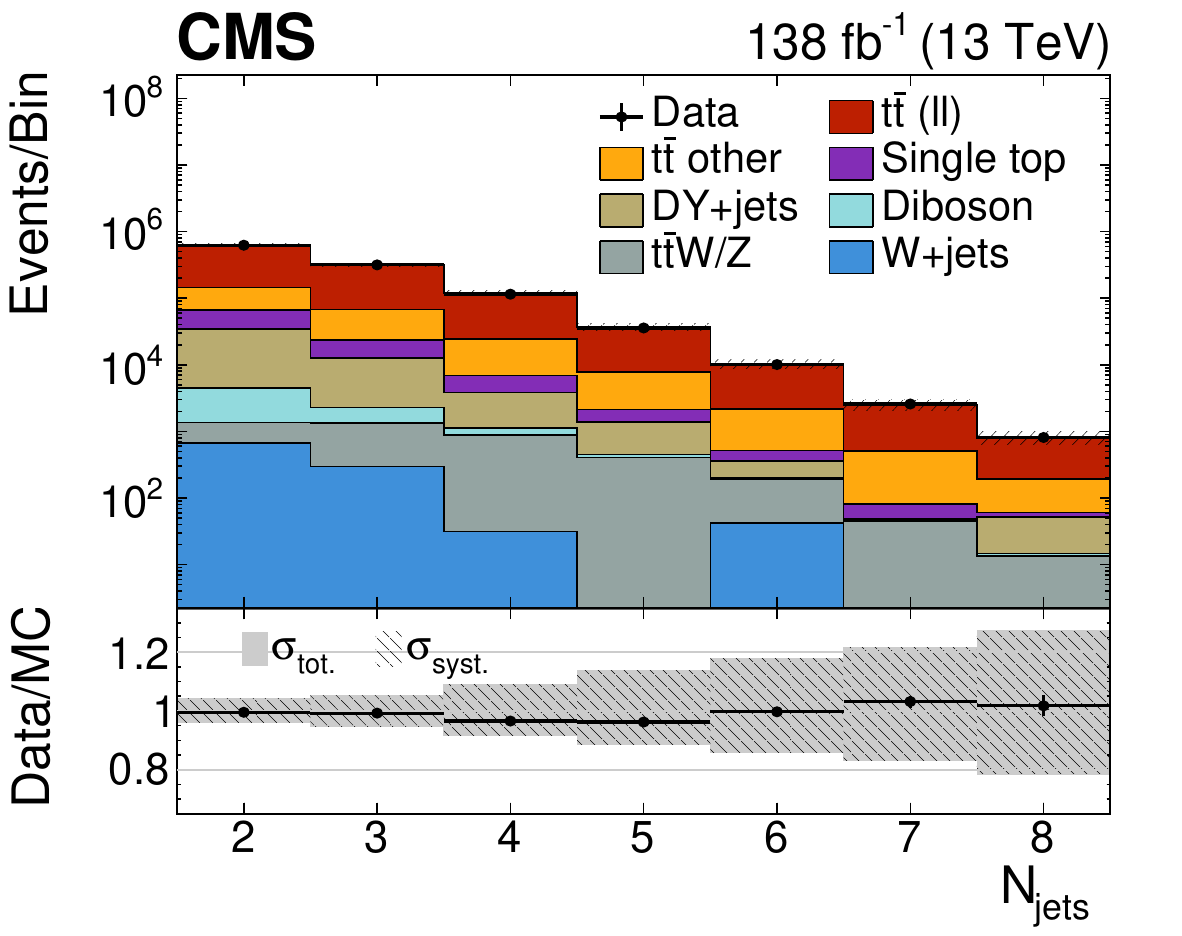}%
\hfill%
\includegraphics[width=0.48\textwidth]{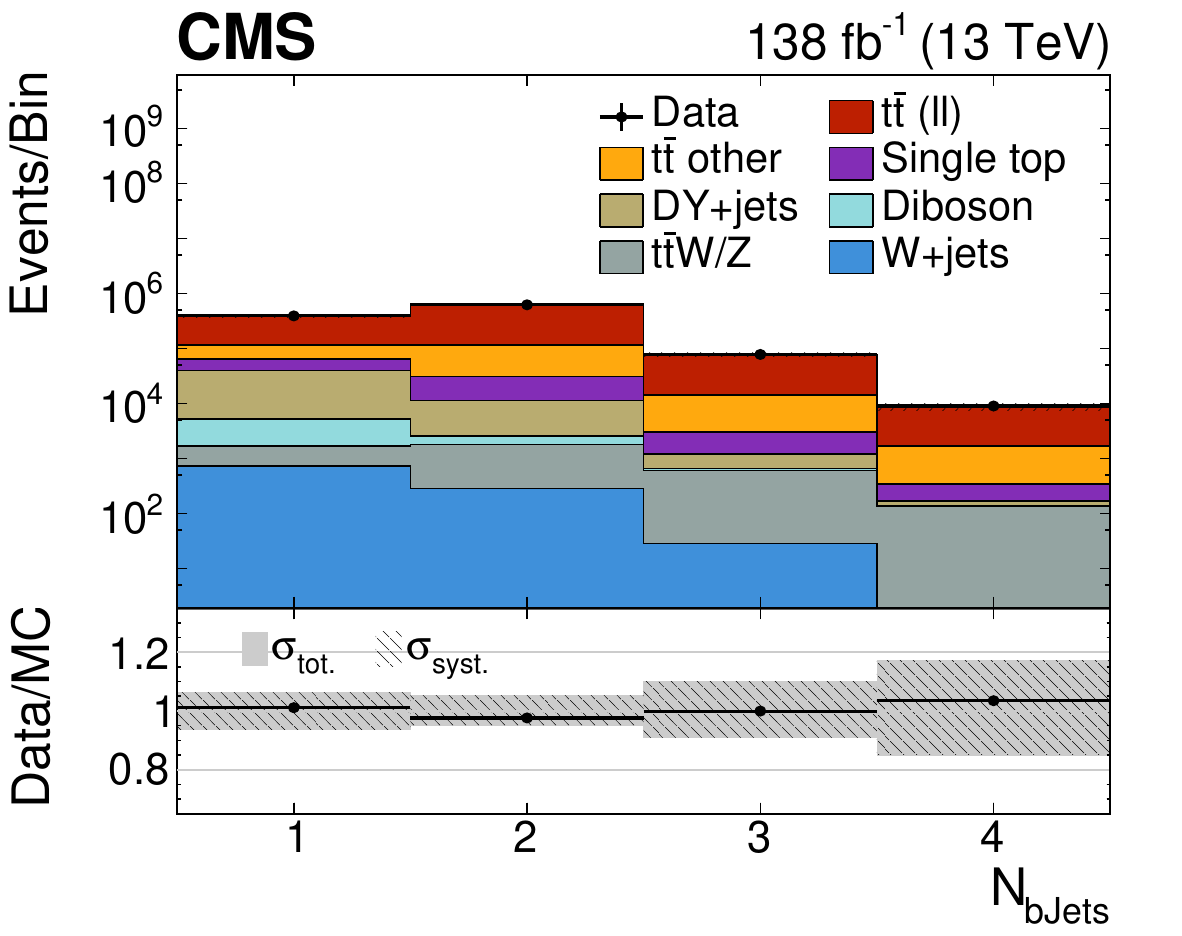}
\caption{Observed (black markers) and expected (filled histograms) distributions of leading lepton \pt (upper left), leading jet \pt (upper right), the number of jets (lower left), and the number of \PQb tagged jets (lower right), after event selection. The hatched (grey) areas denote the systematic (total) uncertainties on the expected yields. Vertical and horizontal error bars on the data points represent the statistical uncertainty and the bin width, respectively. Events from all data-taking periods and all channels are combined. The lower panel of each plot shows the ratio between observed and expected distributions. The last bin includes all events above the plotted range.}
\label{fig:control_all}
\end{figure}

The numbers of expected events from simulation and of observed events in data are shown in Table~\ref{tab:yields_combined}. The signal process \ttbar (\ttll) comprises 78\% of the selected events. The largest background contribution arises from other \ttbar processes, contributing to 13\% of the selected sample. It mostly contains events where at least one \PW~boson decays to a \PGt lepton. The next background contributions in size are single top and DY processes, with 4.2 and 4.0\% contributions to the selected samples, respectively. Distributions of a selected set of kinematic observables are shown in Fig.~\ref{fig:control_all}. The simulated samples reproduce the distributions in data well.

\begin{table}[!htb]
\centering
\topcaption{Data and MC simulation yields after the event selection, combined for all data-taking periods and split by channels. The uncertainties on the expected yields include systematic and statistical uncertainties (discussed in Section~\ref{sec:sysunc}). The relative contribution in percent of each process to the total expected yield of a channel is given in parentheses.}
\label{tab:yields_combined}
\renewcommand{\arraystretch}{1.3}
\begin{tabular}{ccccc}
    & \multicolumn{4}{c}{Yields / $10^3$} \\
    \cline{2-5}
    Process & \ee & \mumu & \emu & All \\ \hline
    \ttbar (\ttll) & 116 (75.2\%) & 221 (74.1\%) & 531 (80.2\%) & 867 (77.9\%) \\
    \ttbar other & 18.2 (11.8\%) & 38.6 (12.9\%) & 90.5 (13.7\%) & 147 (13.2\%) \\
    Single top & 6.15 (4.0\%) & 11.7 (3.9\%) & 28.4 (4.3\%) & 46.3 (4.2\%) \\
    DY+jets & 12.6 (8.2\%) & 24.8 (8.3\%) & 6.87 (1.0\%) & 44.2 (4.0\%) \\
    Diboson & 0.70 (0.5\%) & 1.24 (0.4\%) & 2.43 (0.4\%) & 4.36 (0.4\%) \\
    \ttWZ & 0.49 (0.3\%) & 0.89 (0.3\%) & 1.82 (0.3\%) & 3.20 (0.3\%) \\
    {\PW}+jets & 0.08 (0.0\%) & 0.08 (0.0\%) & 0.88 (0.1\%) & 1.03 (0.1\%) \\[\cmsTabSkip]
    Sum MC & $154\pm10$ & $298\pm14$ & $662\pm27$ & $1114\pm47$ \\
    Data & 151.568 & 302.168 & 649.504 & 1103.240 \\
\end{tabular}
\end{table}

\section{Improving the \texorpdfstring{\ptmiss}{ptmiss} resolution for dileptonic \texorpdfstring{\ttbar}{tt} events}
\label{sec:dnn}

Since the \ptvecmiss is used as a measure of the \ptvec of the two prompt neutrinos (\pTVecnunu) produced in dileptonic \ttbar decays, it is important to
ensure an accurate reconstruction of its magnitude and direction. Poor resolution or large biases of the reconstructed \ptvecmiss, denoted as \ptvecmissrec, can compromise the stability of the unfolding
process and lead to a reduced number of bins for the final measurements. Other than those coming from the two neutrinos from top quark decays, there are two additional sources that contribute to the evaluation of \ptvecmiss in dileptonic \ttbar events. The first source corresponds to the production of nonprompt neutrinos from semileptonic hadron decays in jets, especially in jets originating from \PQb quarks. These nonprompt neutrinos lead to genuine \ptvecmiss and are therefore included in the calculation of the generated \ptmiss of an event. This quantity, defined as the magnitude of the vector sum of all simulated neutrino \ptvec is denoted as \ptvecmissgen. The second source is the mismeasurement of particle momenta during reconstruction, with the largest impact arising from the mismeasurements in jets. A regression based on a DNN, targeting \ptvecmissgen, is derived to correct for the mismeasurements. The method, detailed in the following, corrects for reconstruction effects based on the targeted final state, while being independent of the exact particle kinematic properties and the description of hadronization.

A feed-forward, fully-connected DNN with two output nodes is constructed and trained using the \textsc{TensorFlow}~\cite{Abadi:2016kic} backend with a \textsc{keras}~\cite{chollet2015keras} interface. The DNN architecture consists of two hidden layers with 512 and 256 nodes for the first and second layers, respectively. In order to simultaneously correct the direction and magnitude of \ptvecmiss, the two output nodes of the DNN are the $x$ and $y$ components of the difference between \ptvecmisspuppi and \ptvecmissgen. The output of the DNN is then used to correct the \ptvecmisspuppi. The network is trained based on the alternative \MCATNLO{}+\PYTHIA{} \ttbar sample, which is statistically independent of the \POWHEG{}+\PYTHIA sample used in the unfolding procedure. The training is performed on dileptonic \ttbar events passing the selection detailed in Section~\ref{sec:selection}, but omitting the requirement on \ptmissdnn. The training sample is reweighted so that the \ptmissgen distribution is uniform up to 500\GeV, and a constant weight is applied to events above 500\GeV. This allows the DNN to perform on a wider \ptvecmissgen range rather than being biased to a specific range with the largest population in the nominal training sample. To account for varying data-taking conditions, the training is performed separately for each data-taking period, based on a total of approximately 5 million events.

The DNN is based on 17 input observables, including the $x$ and $y$ components of \ptvecmisspuppi, \ptvecmisspf, the \ptvec sum of all jets, the \ptvec of the leading and subleading jets, and the \ptvec of the leading lepton. Additional inputs include the invariant mass of the two leptons and all jets, the invariant mass of the two leading jets, the energy of the leading jet, and the scalar \pt sum of all jets. It is confirmed that all inputs are well modeled in the simulation by performing binned profile-likelihood fits of each input distribution to data. All systematic uncertainties (discussed in Section~\ref{sec:sysunc}) are included in the fit as nuisance parameters. In each case, the goodness of fit is evaluated based on a saturated model~\cite{Cousins2013GeneralizationOC}, as implemented in the CMS statistical analysis tool, \textsc{combine}~\cite{CMS:2024onh}. The procedure is repeated for every 2D distribution of two inputs to ensure a correct description of correlations between inputs by the simulation. Furter details on the goodness-of-fit tests can be found in Ref.~\cite{MeuserPhD}.

\begin{figure}[p!h]
\centering
\includegraphics[width=0.8\textwidth]{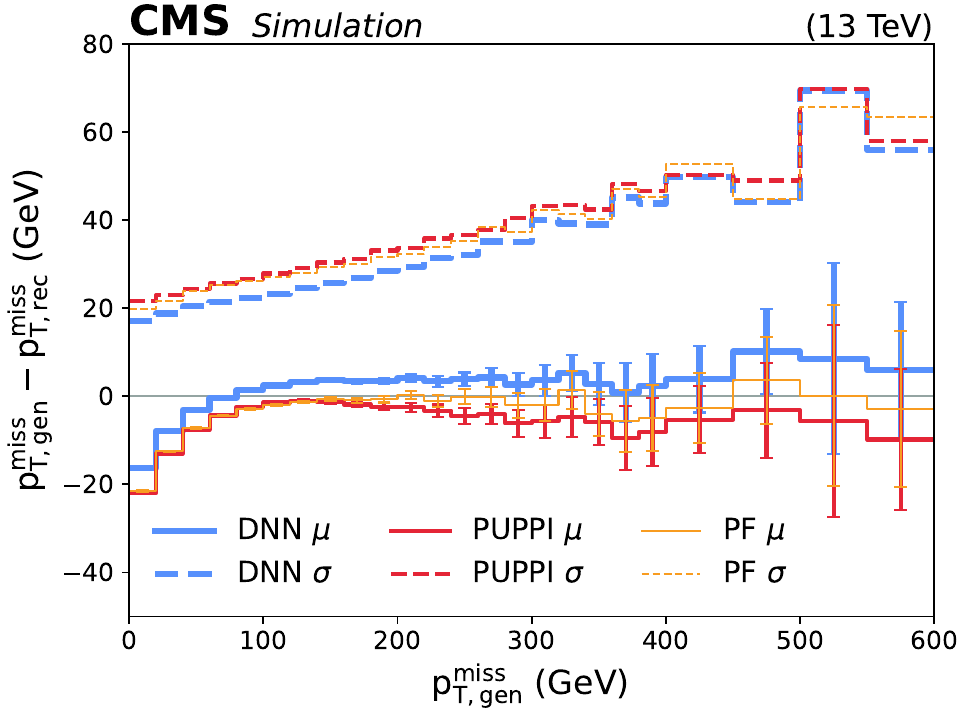} \\
\includegraphics[width=0.8\textwidth]{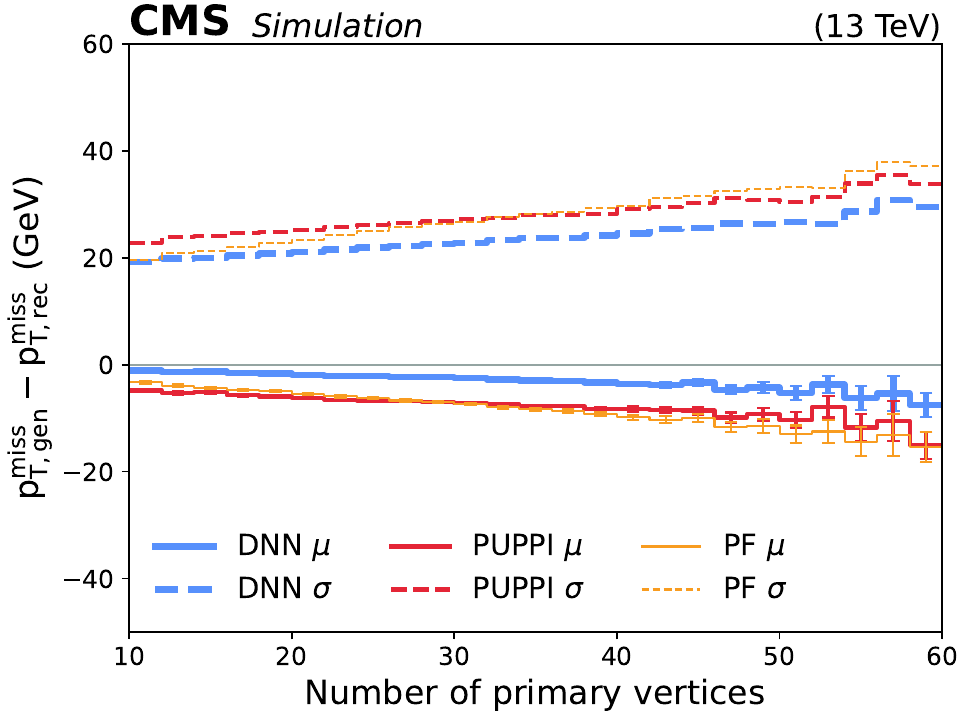}
\caption{Difference between \ptmissgen and \ptmissrec as a function of the \ptmissgen (upper) and the number of primary vertices (lower) for signal events. The mean difference between the generated and the \ptmissrec per bin is shown as solid line, while the dashed line shows the standard deviation $\sigma$, which corresponds to the resolution. The results for \ptmiss corrected by the DNN regression (thick light blue), derived with the PUPPI algorithm (thin red), and the PF algorithm (thinner orange) are shown. A simulated sample for the 2018 data-taking period is used.}
\label{fig:dnn_performance}
\end{figure}

The performance of \ptmissdnn is compared to \ptmisspf and \ptmisspuppi in Fig.~\ref{fig:dnn_performance}, using a simulated sample corresponding to the 2018 data-taking period. The mean ($\mu$) and the resolution (standard deviation, $\sigma$) of the difference between \ptmissgen and \ptmissrec is shown as a function of the \ptmissgen and the number of PVs. As a function of both observables, \ptmissdnn improves on the resolution, and inclusively it shifts the mean closer to zero in the phase space of this analysis compared to the other \ptmiss reconstruction algorithms. The negative bias at low \ptmissgen arises from the finite \ptmiss resolution which causes the \ptmissrec to be on average larger than \ptmissgen in events with genuinely small \ptmissgen~\cite{CMS:2019ctu}. The DNN can marginally correct for this effect thanks to a better \ptmiss resolution. Overall, the resolution of \ptmiss is improved by approximately 15\% by the DNN with respect to \ptmisspuppi, while the resolution of $\phi(\ptvecmiss)$ is improved by around 12\%. The improvements allow for a finer binning in the differential measurements of the target observables as they result in a reduction of bin-to-bin migration while ensuring a stable unfolding.

\section{Systematic uncertainties}
\label{sec:sysunc}

The differential cross section measurements are affected by various sources of systematic uncertainty. They can be grouped into two categories, namely experimental and theoretical uncertainties. The impact of each source is evaluated by varying the uncertainty source by $\pm$1 standard deviation or by using alternative simulation settings. These variations can affect the response matrix used in the unfolding procedure, as well as the background estimation. For each systematic variation the cross sections are extracted and differences with the nominal values are taken as uncertainties. Because of different detector and/or simulation conditions, dedicated correlation models between different data-taking periods are employed for each source of systematic uncertainty. Unless stated otherwise, the uncertainty is assumed to be fully correlated between different years of data taking.

Experimental uncertainties include those originating from jet reconstruction, corrections applied to simulated samples to improve the shape agreement of data and simulation, normalization of background processes, integrated luminosity measurement, and determination of unclustered energy. The impact from the jet energy scale (JES) uncertainty is estimated by shifting all jet momenta according to seven independent uncertainty sources, which have different treatments of correlations between the data-taking periods. The corresponding shifts in the jet momentum are propagated to \ptmiss~\cite{CMS:2016lmd,CMS:2019ctu}. The uncertainty on the jet energy resolution (JER) is introduced when smearing jet energies in simulation to match the resolution observed in data. To evaluate its impact, the smearing is repeated with the JER shifted within its uncertainty. This uncertainty is assumed to be uncorrelated between the data-taking periods.

Correction factors are applied to simulation to account for the differences with data when describing \PQb tagging, lepton selection, lepton energy scale, identification of jets from pileup, trigger selection, and ECAL first level trigger inefficiencies~\cite{CMS:2024ppo}. Each of these correction factors is varied within its uncertainty to estimate its impact on the measured cross section. The statistical component of the uncertainty related to the muon identification efficiency, as well as trigger efficiencies, is considered to be uncorrelated between the data-taking periods. Variations in lepton energy scale are propagated to \ptmiss. In addition, the corrections to \PQb tagging efficiencies are re-evaluated separately for systematic variations in JES, JER, identification of jets from pileup, and most of the theoretical uncertainties.

The uncertainties on the integrated luminosities are 1.2, 2.3, and 2.5\%, for 2016, 2017, and 2018, respectively, with partial correlations between data-taking periods. This results in a combined uncertainty of 1.6\%~\cite{CMS:2021xjt,CMS:2018elu,CMS:2019jhq}.

The distribution of additional \pp collisions in simulation is reweighted according to the expected distribution in data assuming a total inelastic \pp cross section of 69.2\unit{mb} \cite{CMS:2018mlc}. The reweighting is repeated shifting the inelastic \pp cross section by $\pm$4.6\% to derive the corresponding uncertainty.

The normalization of all background processes directly impacts the cross section measurements. The magnitudes of the corresponding uncertainties applied to the normalization are taken from Ref.~\cite{CMS:2024ybg}. The largest background contribution, denoted as \ttbar other, is assigned a 5\% normalization uncertainty. For single top and DY+jets production, a 20\% and a 30\% normalization uncertainty is assigned, respectively. Subdominant background processes, such as diboson production, {\PW}+jets, and \ttWZ production are assumed to have a 30\% normalization uncertainty. Each background normalization is varied individually within the given uncertainty. The simulated samples are normalized to the integrated luminosities of the corresponding data-taking periods~\cite{CMS:2021xjt,CMS:2018elu,CMS:2019jhq}. The sample normalization is shifted according to the integrated luminosity uncertainties, which are split into uncorrelated and correlated components between the different data-taking periods.

The \ptmiss measurement is affected by energy deposits that remain unclustered. To estimate their impact, the energy from charged and neutral hadrons, as well as photons and hadrons reconstructed in the $\eta$ region covered by the forward hadron calorimeter, is shifted according to their energy resolutions and propagated to \ptmiss. The resulting uncertainty is assumed to be uncorrelated between data-taking periods.

Theoretical uncertainty sources include variations that affect all simulated samples, as well as variations that are only applied to the \ttbar simulation. The impact of initial-state radiation and final-state radiation is estimated by shifting the renormalisation scale for QCD emission in the parton shower by a factor of two and a factor of one half, separately for both initial-state and final-state radiation scales. The uncertainty on the choice of renormalization and factorization scales is estimated in a similar way, shifting the scales up and down by a factor of two, both individually and simultaneously, excluding the configuration where the two scales are shifted simultaneously in opposite directions. The uncertainty on the cross section results is given by the envelope of the corresponding shift variations in the measurements. Uncertainties due to the matching of ME and PS calculations are taken into account by reweighting the nominal \ttbar samples using weights derived with the DNNs using the classification for tuning and reweighting approach~\cite{Andreassen:2019nnm, CMS-PAS-MLG-24-001} to shift the $h_{\text{damp}}$ parameter of the \POWHEG event generator. This parameter controls the merging between ME and PS and regulates the high \pt radiation. The impact of the choice of the scheme to handle the overlap between \tW and \ttbar production~\cite{Re:2010bp} is estimated by taking the difference between the \tW sample based on DR and DS as additional uncertainty into account.

The remaining theoretical uncertainties are only taken into account for the \ttbar simulation. Alternative \ttbar samples are used to estimate the impact of varying the top quark mass by 1\GeV, using different tunes for the underlying event modeling as described in Ref.~\cite{CMS:2019csb}, employing two alternative approaches for color reconnection, and enabling early resonance decays~\cite{CMS:2022awf}. The energy measurement of jets from \PQb quarks is impacted by semileptonic decays of \PQb hadrons. The assumed branching fractions, $\mathcal{B}$, taken from Ref.~\cite{ParticleDataGroup:2024cfk}, are shifted within the reported uncertainty to estimate the effect on the cross section measurements. The impact of reweighting the top quark \pt spectra to NNLO accuracy is evaluated in the cross section measurement by repeating the measurement without the reweighting and taking the difference as a one-sided uncertainty.

Uncertainties associated with the PDF description are evaluated by reweighting the simulated \ttbar sample using weights based on 100 eigenvectors of the NNPDF3.1 Hessian error PDF set~\cite{Ball:2011uy,NNPDF:2017mvq}, keeping the overall normalization fixed. The envelope of all 100 variation shifts is taken as a systematic uncertainty.

\section{Results}
\label{sec:results}

\begin{figure}[htp!]
\centering
\includegraphics[width=0.48\textwidth]{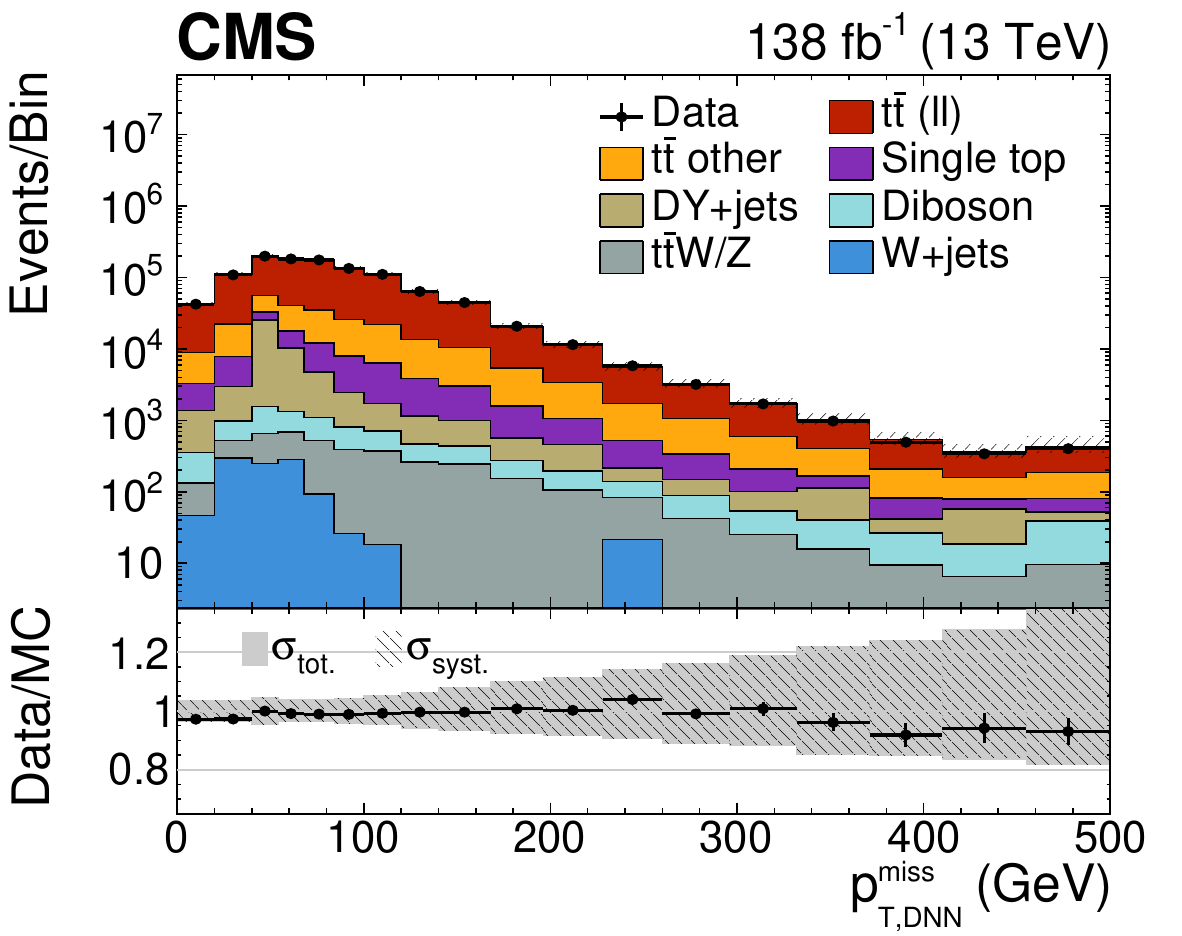}%
\hfill%
\includegraphics[width=0.48\textwidth]{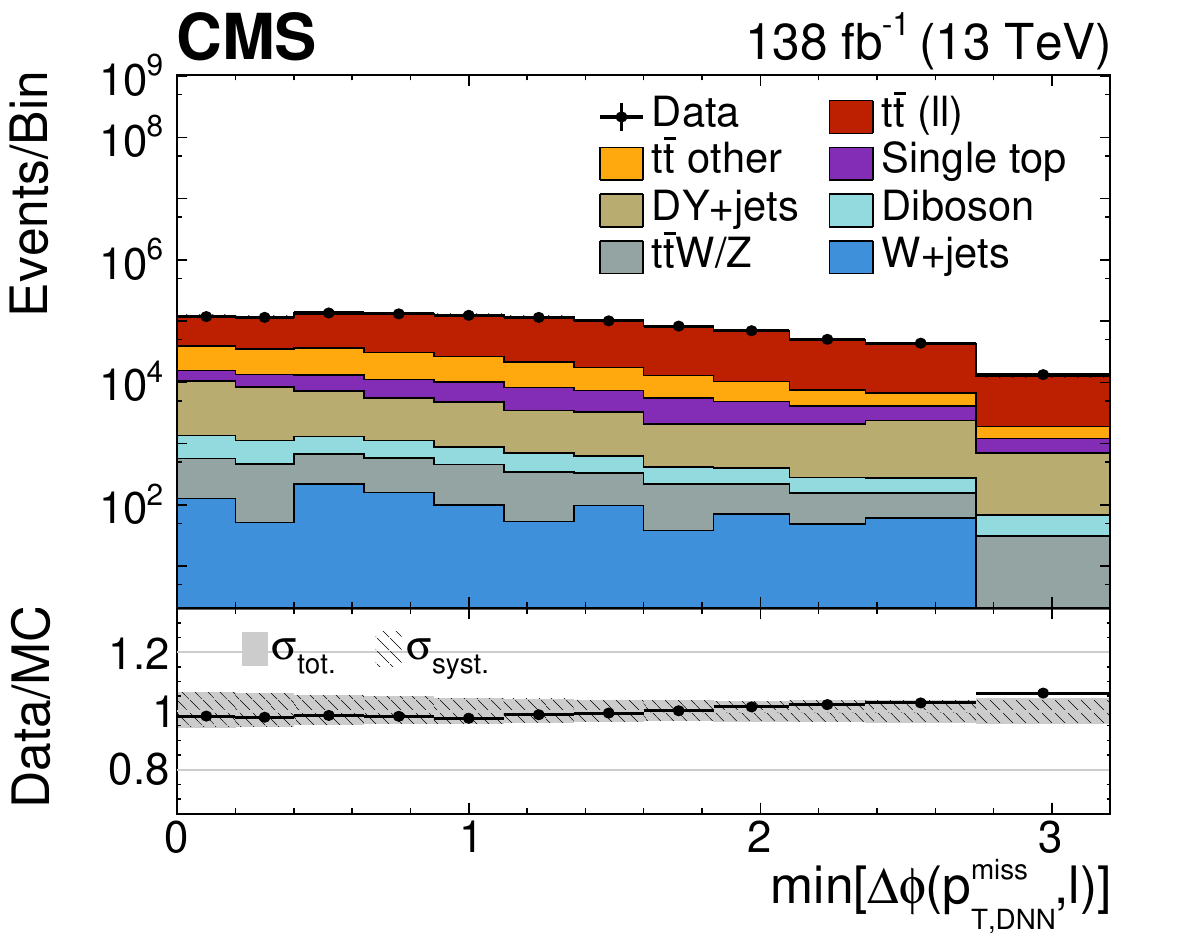} \\
\includegraphics[width=0.77\textwidth]{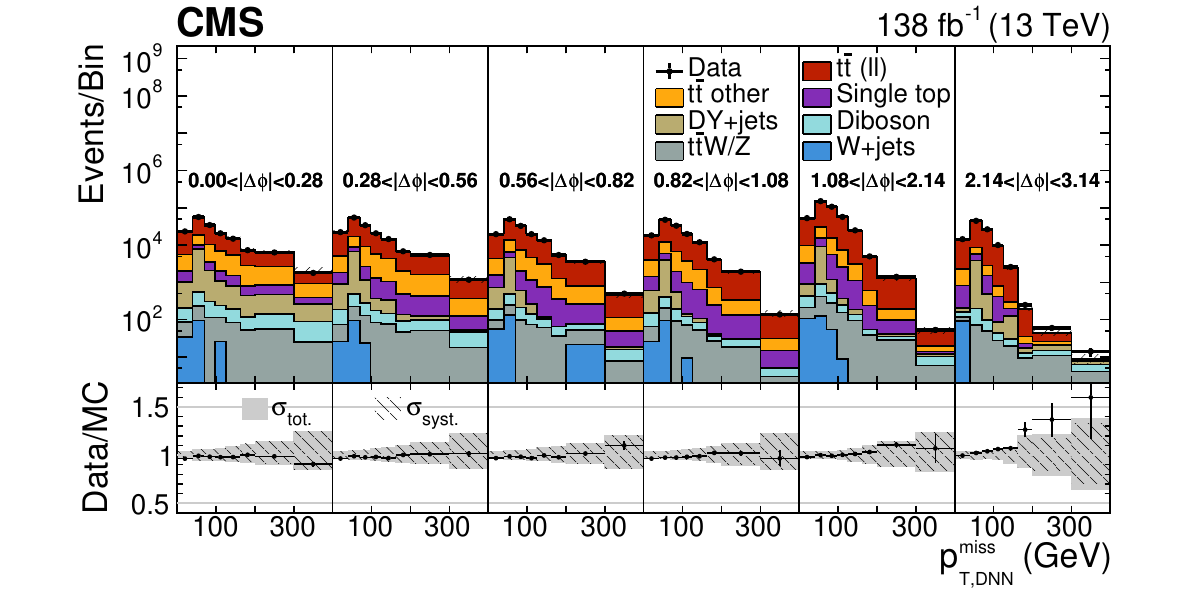}
\caption{The observed (black markers) and simulated distributions of \ptmissdnn (upper left), \phiNextLepMETVec (upper right), and the 2D distribution (lower) with both observables are shown. Events from all data-taking periods and all channels are combined. The hatched (grey) areas indicate the systematic (total) uncertainty on the simulation. Vertical and horizontal error bars on the data points represent the statistical uncertainty and the bin width, respectively. The lower panel of each plot shows the ratio of observed data to the expectation from MC simulation in each bin. The last bin includes all events above the plotted range.}
\label{fig:result}
\end{figure}

The detector-level distributions of \ptmissdnn, \phiNextLepMETVec, and the 2D distribution of both observables are compared to the expectation from simulation in Fig.~\ref{fig:result}. In general, good agreement between the observed data and the simulation is found. For the 2D distribution, a slight over fluctuation at large \phiNextLepMETVec and \ptmissdnn is found in data, still compatible with simulation considering the large statistical uncertainties in this phase space.

\begin{table}[tp!]
\centering
\topcaption{Definition of the fiducial phase space for the same-flavor and the different-flavor channels.}
\label{tab:fiducial}
\renewcommand{\arraystretch}{1.2}
\begin{tabular}{ccc}
    Parameter & \ee{}/\mumu & \emu \\ \hline
    Lepton \pt & \multicolumn{2}{c}{$>$20\GeV} \\
    Lepton \abseta & \multicolumn{2}{c}{$<$2.4} \\
    Number of leptons & \multicolumn{2}{c}{$=$2} \\
    Invariant mass of the two leptons & \multicolumn{2}{c}{$>$20\GeV} \\[\cmsTabSkip]
    Jet \pt & \multicolumn{2}{c}{$>$30\GeV} \\
    Jet \abseta & \multicolumn{2}{c}{$<$2.4} \\
    $\DR$(Jet,Lepton) & \multicolumn{2}{c}{$>$0.4} \\
    Number of \PQb jets & \multicolumn{2}{c}{$\geq$2} \\[\cmsTabSkip]
    Dineutrino transverse momentum (\pTnunu) & $>$40\GeV & \NA \\
\end{tabular}
\end{table}

The differential cross sections as functions of \pTnunu and \phiNextLepNuNuVec are derived at particle-level by unfolding the detector-level distributions of \ptmissdnn and \phiNextLepMETVec in a fiducial phase space, as described below. The definition of the fiducial phase space summarised in Table~\ref{tab:fiducial} follows closely the definitions detailed in Ref.~\cite{CMS:2024ybg}, with the additional requirement of $\pTnunu>40\GeV$ in the same-flavor channels. The momenta of photons within a cone of $\DR<0.1$ around the lepton are added to the lepton momentum, accounting for bremsstrahlung radiations. Particle-level jets containing a B hadron are identified as originating from a b quark. For a better comparison of the measurement with fixed-order predictions, non-prompt neutrinos originating from hadron decays are included in the jet clustering, as recommended in Ref.~\cite{Czakon:2020qbd}. The binning applied in the final measurements is optimized taking into account the detector resolution, migration effects, and statistical uncertainties. The differential cross section results are extracted from the observed distributions in two steps. In the first step, the estimated contributions from all backgrounds except for \ttbar other are subtracted from the observed distributions. In the second step, the observed distributions are unfolded to the fiducial phase space. Within the second step, the contributions from \ttbar other are removed by applying a fixed signal fraction per bin, derived based on the ratio of \ttbar signal events to all selected \ttbar events.

\begin{figure}[htp!]
\centering
\includegraphics[width=0.65\textwidth]{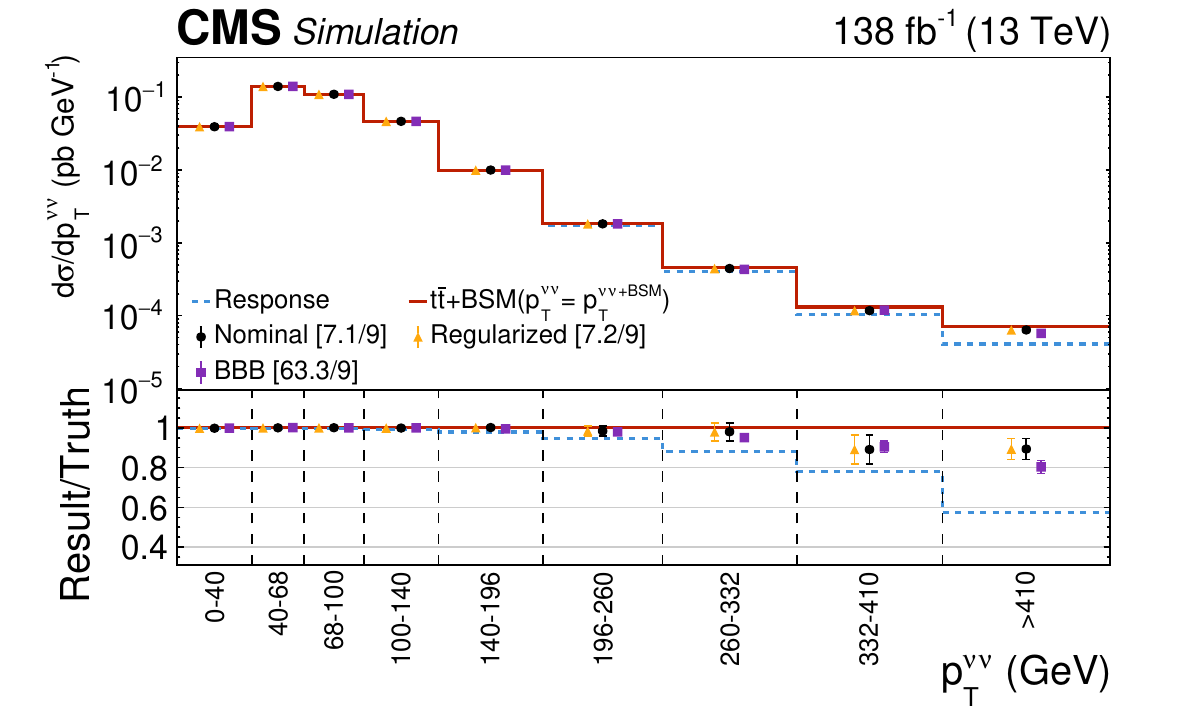} \\
\includegraphics[width=0.65\textwidth]{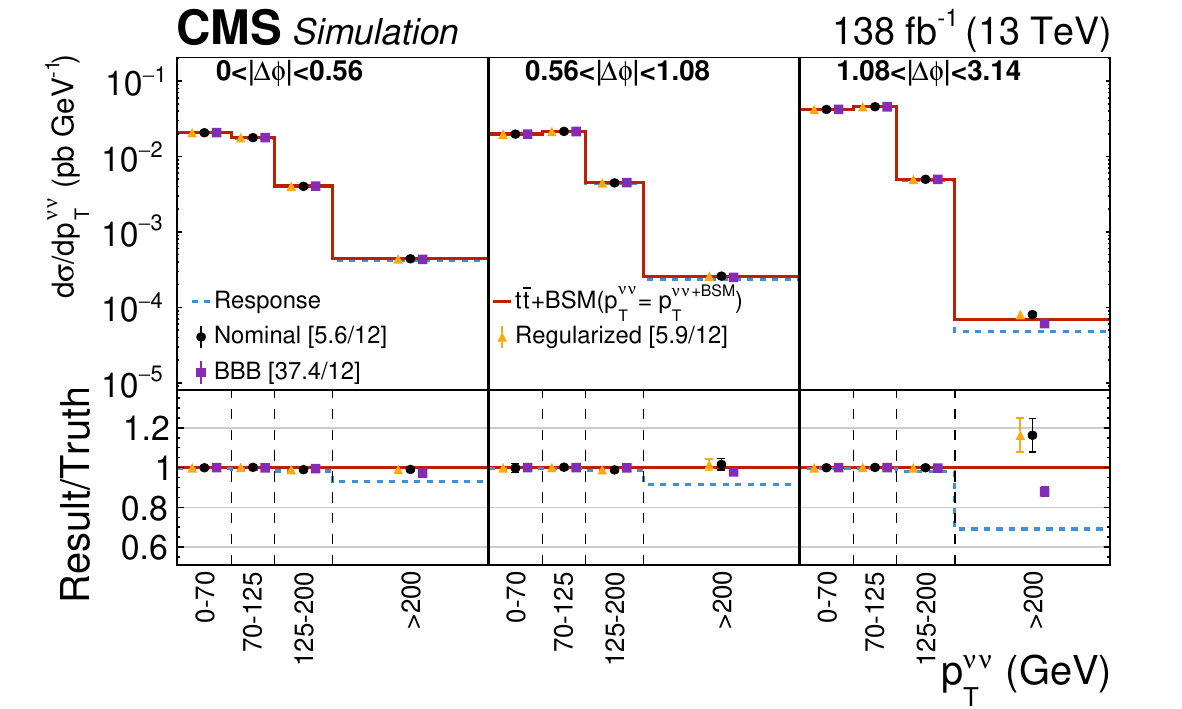}
\caption{Result of the closure test based on simulation accounting for potential BSM contributions based on a top squark pair production scenario with a stop mass of 525\GeV and a neutralino mass of 350\GeV. The potential BSM contribution is scaled by a factor of ten. The test is performed for the \pTnunu (upper) and 2D measurement (lower) using the nominal (black circles), the regularized (orange triangles), and the bin-by-bin unfolding (purple squares), based on all data-taking periods combined. The unfolded distributions are compared to the expected distribution (red solid line) based on the sum of the nominal dileptonic \ttbar signal sample and BSM signal with the corresponding {\chisq}/(number of degrees of freedom) values given in the legend in square brackets. The distribution used for the response matrix is shown in blue (dotted line). The error bars correspond to the statistical uncertainty. The lower part shows the ratios between the unfolded and the expected distributions.}
\label{fig:closure}
\end{figure}

For the unfolding procedure, the \textsc{TUnfold} algorithm~\cite{Schmitt:2012kp} is used, performing a least squares minimization without regularization. The choice of the method is motivated by dedicated unfolding studies, where distorted particle-level distributions as well as pseudodata based on the alternative \MCATNLO{}+\PYTHIA{} \ttbar sample were unfolded using the nominal response matrix. The selected method showed the most stable performance in comparison with a simple bin-by-bin (BBB) unfolding where only the diagonal elements of the response matrix are taken into account as well as with respect to a least squares minimization with regularization. In addition, closure tests accounting for potential BSM contributions based on the top squark pair production scenario shown in Fig.~\ref{fig:diagrams} were performed, for a top squark mass of 525\GeV and a neutralino mass of 350\GeV. Within these tests, pseudodata based on the sum of the nominal \ttbar signal prediction and the prediction for the top squark pair production scenario scaled up by an additional factor of ten were unfolded using the nominal response matrix. The corresponding results for the \pTnunu and the 2D distribution can be found in Fig.~\ref{fig:closure}. Based on the nominal unfolding approach, the expected distributions, where both neutralinos are included in the particle-level definition, can be correctly reproduced. This result confirms that the measurement reproduces deviations from the distribution assumed in the MC simulation, which was used in the construction of the response matrix. The normalized differential cross sections are also derived, with the measured cross section in each bin divided by the total fiducial cross section, allowing for the reduction of several systematic uncertainties.

\begin{figure}[!htb]
\centering
\includegraphics[width=0.48\textwidth]{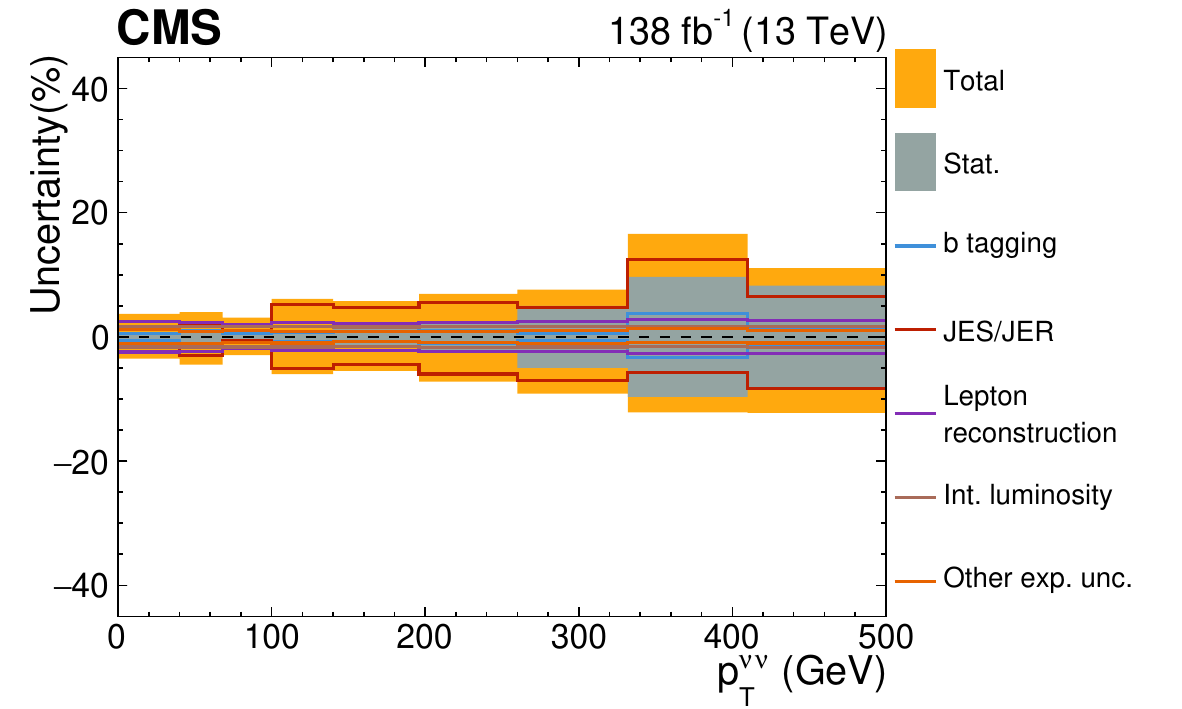}%
\hfill%
\includegraphics[width=0.48\textwidth]{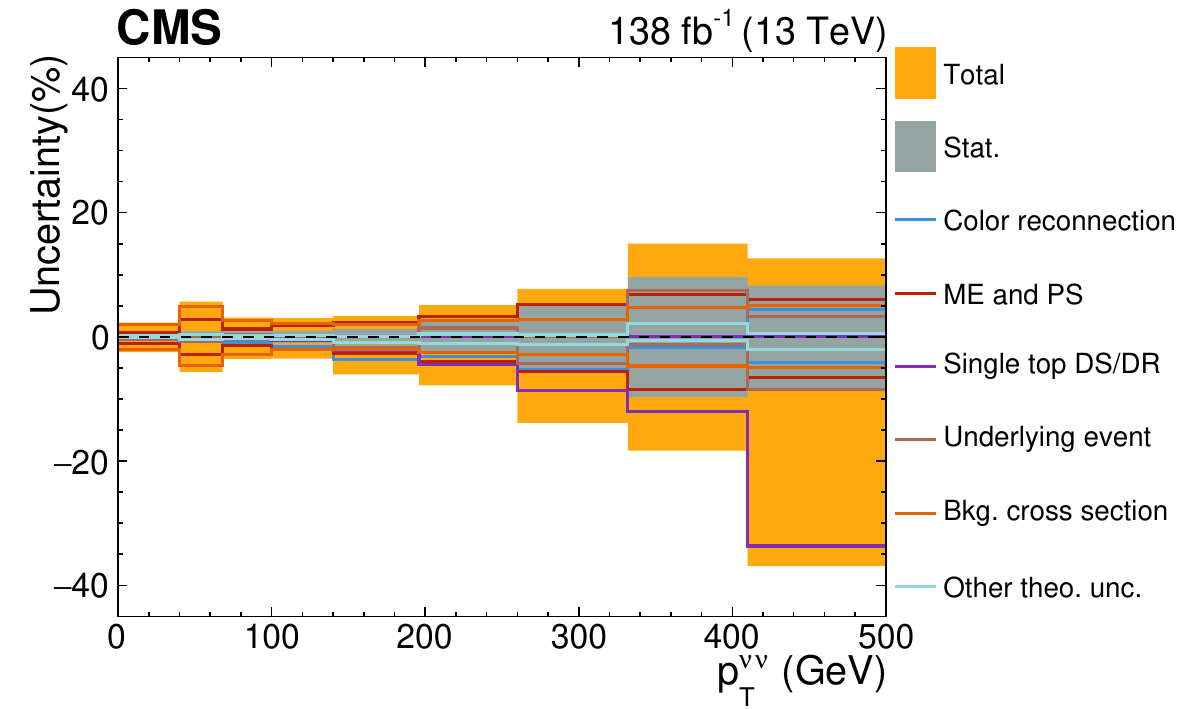} \\[1ex]
\includegraphics[width=0.48\textwidth]{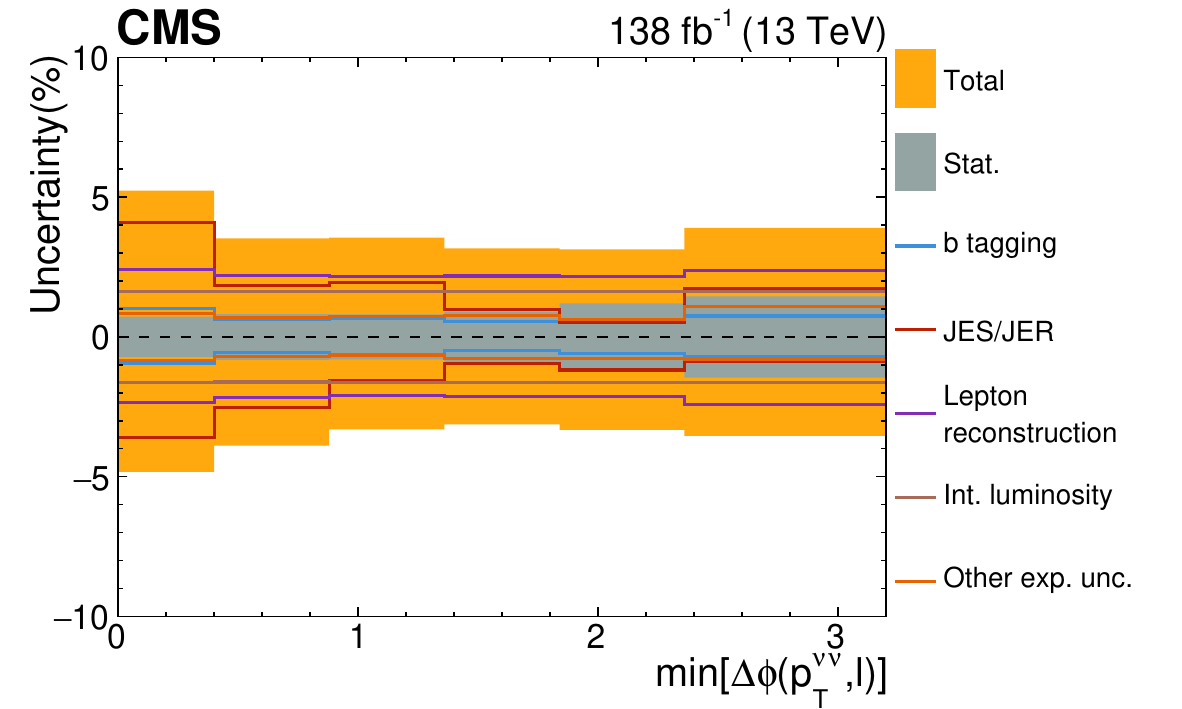}%
\hfill%
\includegraphics[width=0.48\textwidth]{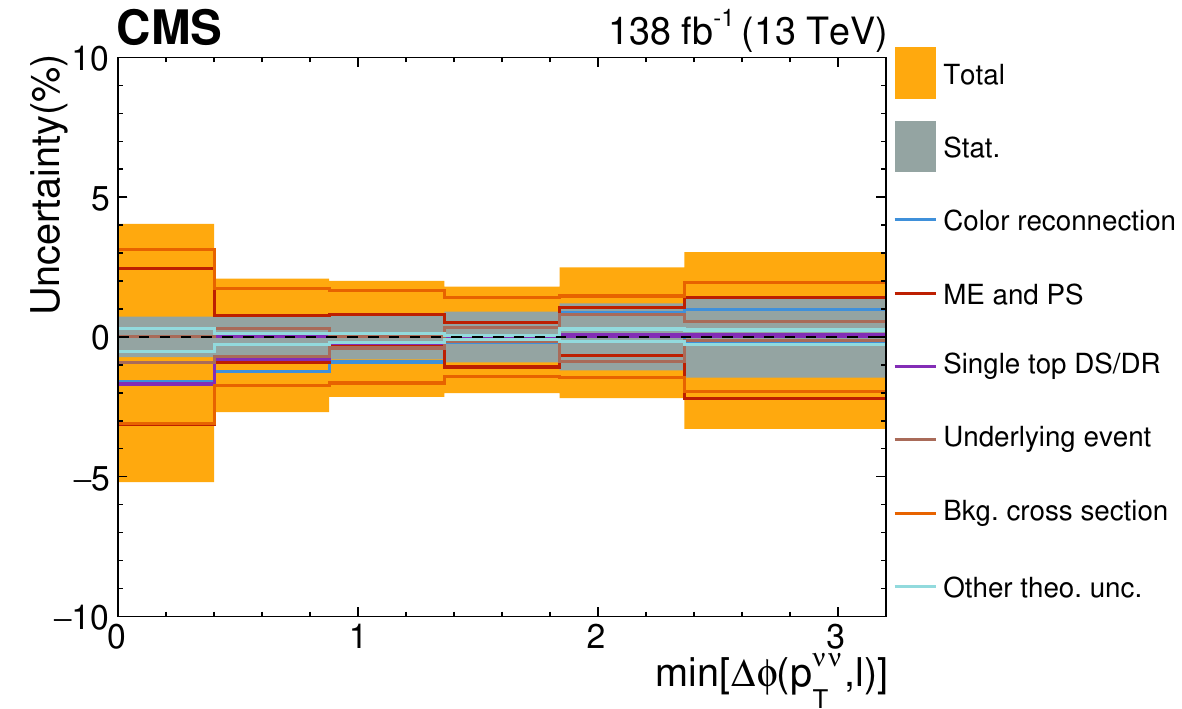} \\[1ex]
\includegraphics[width=0.48\textwidth]{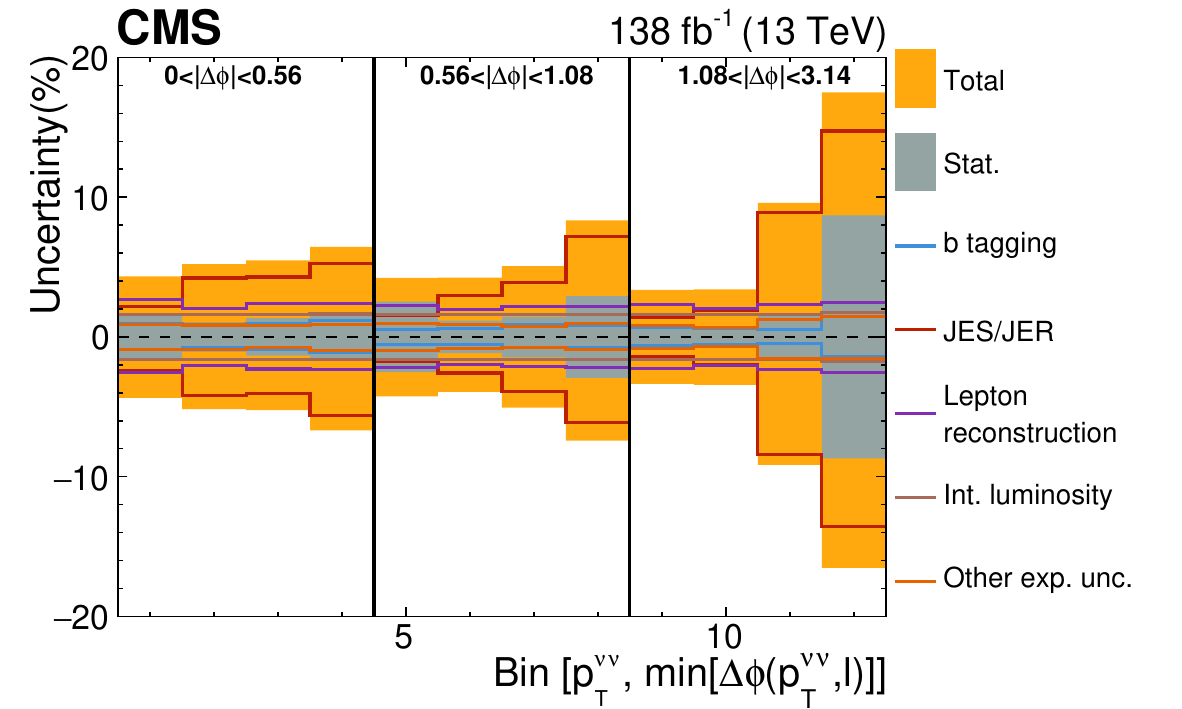}%
\hfill%
\includegraphics[width=0.48\textwidth]{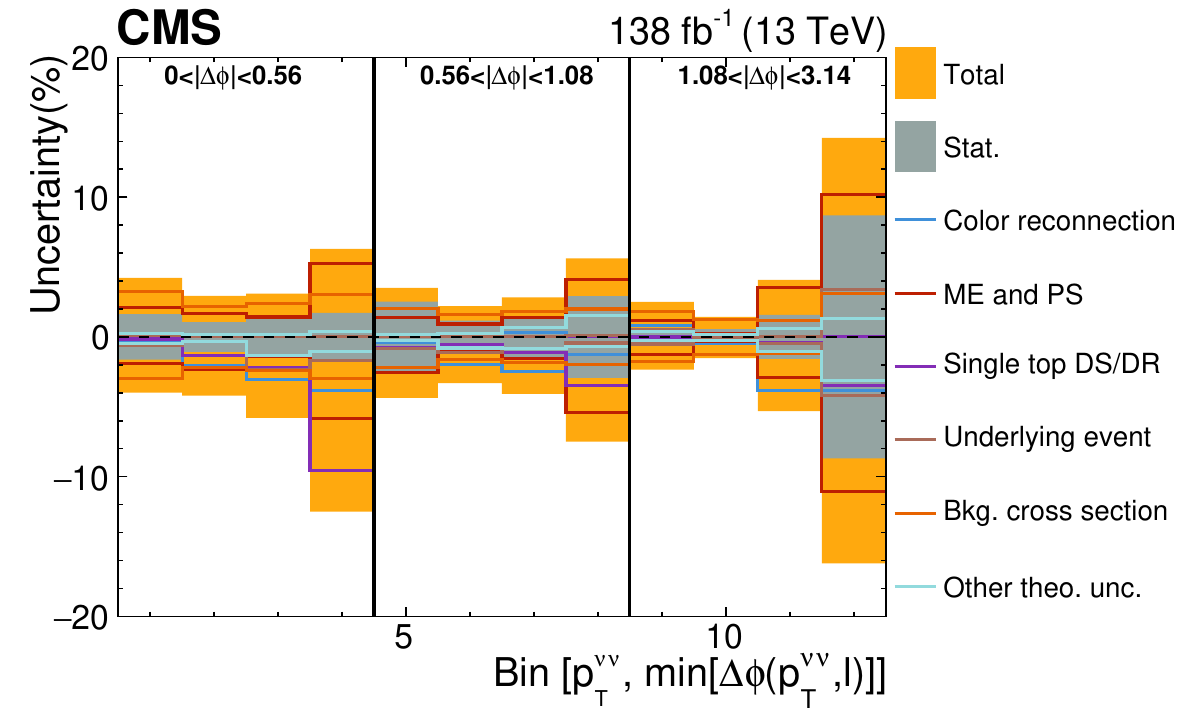}
\caption{Breakdown of the relative uncertainties from experimental (left) and theoretical (right) sources on the differential cross section measurements as functions of \pTnunu (upper), \phiNextLepNuNuVec (middle), and both observables (lower). In the last case, each group of four bins corresponds to \pTnunu bins, and \phiNextLepNuNuVec bin edges are indicated by vertical solid lines. The statistical uncertainty (dark grey) takes into account the size of available event counts in both simulated samples and the recorded data. The last bin includes all events above the plotted range.}
\label{fig:systUncBreakdown}
\end{figure}

The impact of the systematic uncertainties on the cross section measurements is shown in Fig.~\ref{fig:systUncBreakdown}. For the \pTnunu measurement, JES is the dominant experimental uncertainty for most of the bins. At large \pTnunu, the choice of the {\tW}--\ttbar overlap removal scheme is the largest uncertainty contribution, followed by the normalization of the single top production background and the ME scale, which also mainly affects the single top background estimation. For the \phiNextLepNuNuVec measurement, the JES uncertainty is dominant at low values, while at large distances the uncertainty related to the lepton reconstruction has the largest contribution. The 2D measurement shows similar behavior to the \pTnunu measurement, \ie, for most bins the uncertainty associated with JES has the largest contribution. At large \pTnunu in the lowest \phiNextLepNuNuVec bin, the choice of the {\tW}--\ttbar overlap removal scheme is the dominant uncertainty. For the bin with the highest \pTnunu and \phiNextLepNuNuVec, sizeable contributions also arise from the ME-PS matching and ME scale uncertainties.

The bin-to-bin correlations between different sources of uncertainties are extracted from the covariance matrices that include separate terms for statistical and systematic contributions. The statistical component, encoding uncertainties after the unfolding, may contain off-diagonal elements originating from correlations introduced in the unfolding procedure. The same matrices are used to calculate \chisq values when comparing predictions to the measured results. The procedure of constructing the covariance matrices is described in detail in Appendix~A of Ref.~\cite{CMS:2024ybg}.

The measured differential cross sections are compared to five theoretical predictions, namely predictions from the \POWHEG{}+\PYTHIA, \POWHEG{}+\HERWIG, and \MCATNLO{}+\PYTHIA{} \ttbar samples, and fixed-order calculations~\cite{Czakon:2020qbd} at NLO and NNLO accuracy in QCD. The fixed-order calculations use the narrow-width approximation for the top quark decays and include ME scale uncertainties and, in case of the NNLO prediction, also PDF uncertainties. For the three MC-based predictions, a normalization uncertainty of 5\% as well as ME scale uncertainties are taken into account. For the NLO MC simulations, only the acceptance and efficiency effects for the fiducial volume are considered. More details on the cross section extraction can be found in Ref.~\cite{CMS:2024ybg}, which only differs in the choice of regularization.

\begin{figure}[p!h]
\centering
\includegraphics[width=0.65\textwidth]{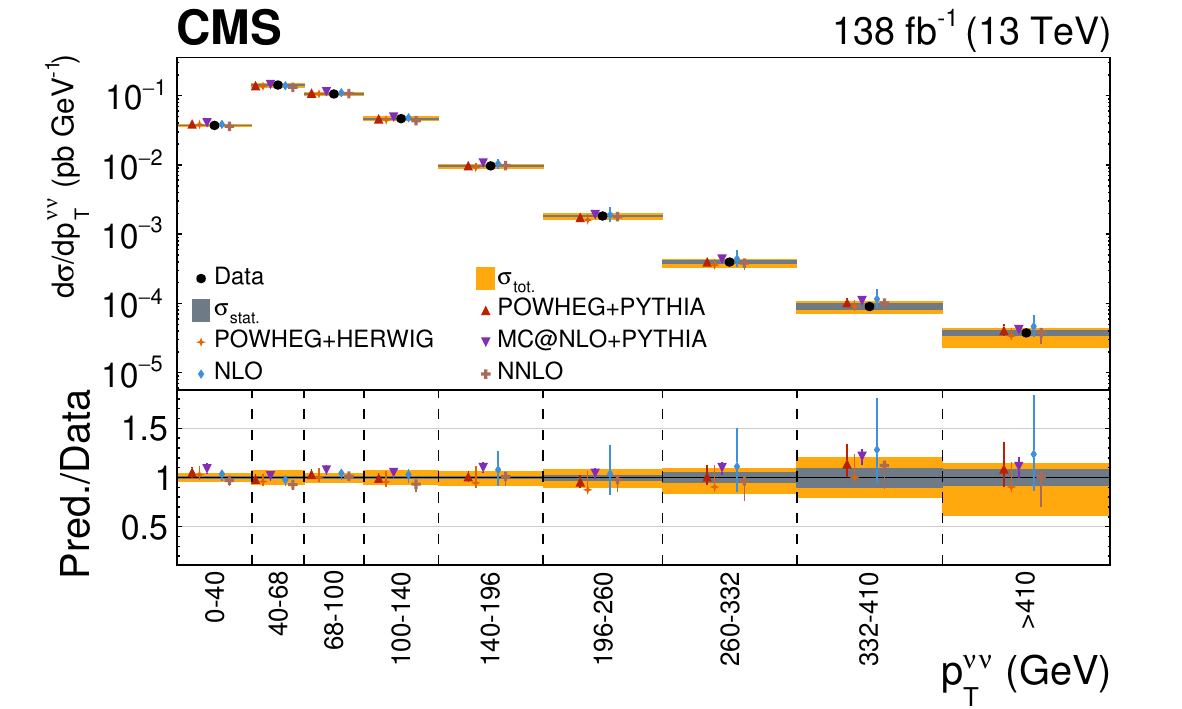} \\
\includegraphics[width=0.65\textwidth]{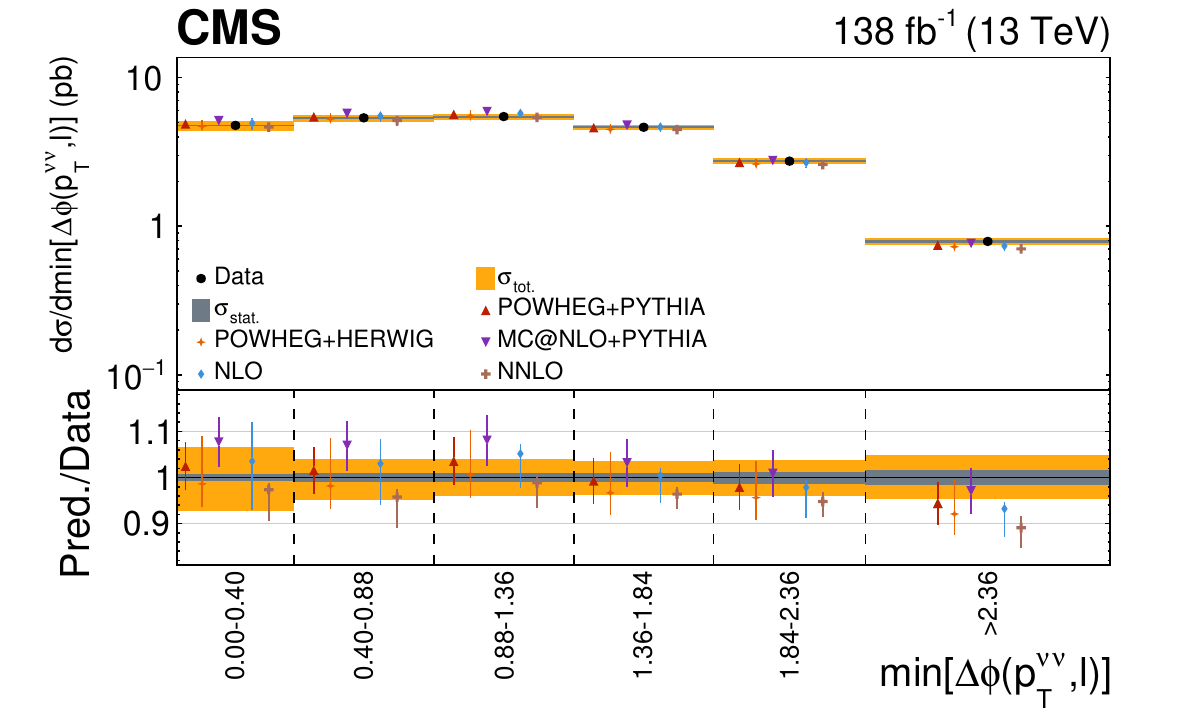} \\
\includegraphics[width=0.65\textwidth]{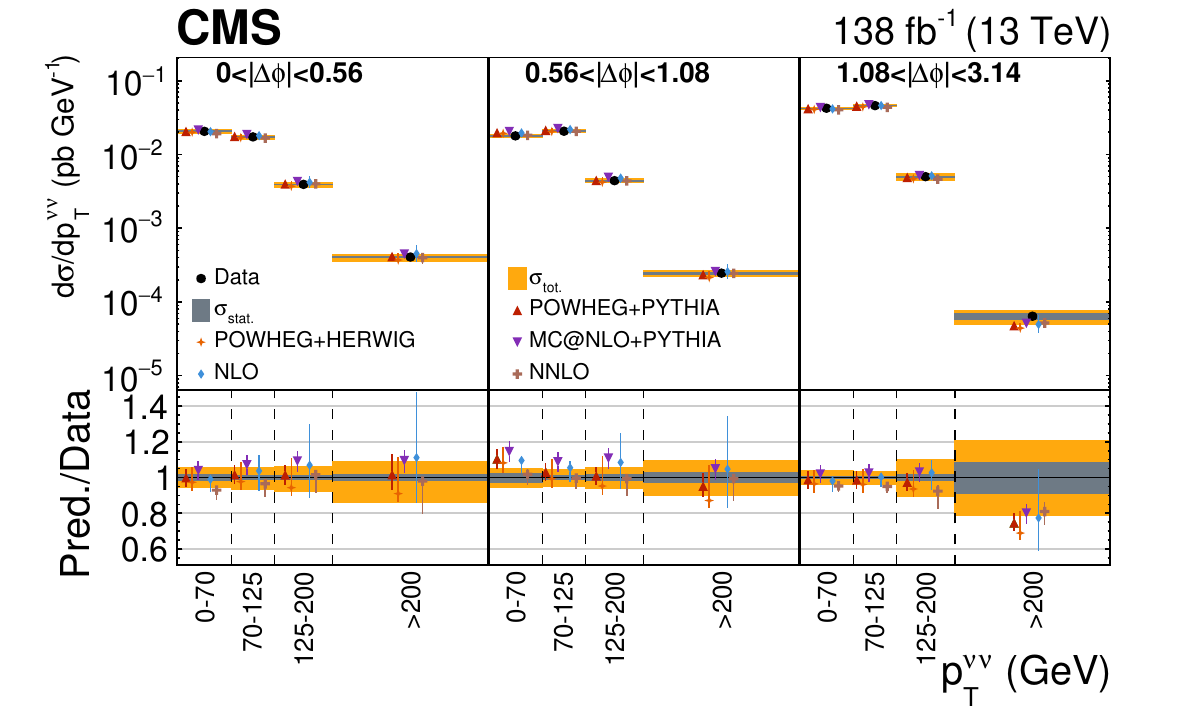}
\caption{The measured differential signal cross sections (black markers) as functions of \pTnunu (upper), \phiNextLepNuNuVec (middle), and double-differential in both observables (lower) are shown. The theoretical predictions from \POWHEG{}+\PYTHIA (dark red), \POWHEG{}+\HERWIG (orange), \MCATNLO{}+\PYTHIA (purple), and the fixed-order NLO (light blue) and NNLO (brown) calculations are compared to the measurement. The cross sections in the last \pTnunu bin of the upper (lower) plot are determined for $\pTnunu>410$ (200)\GeV and normalised to a bin width of 90 (200)\GeV for visualisation purposes. The total (statistical) uncertainty on the measurement is shown as an orange (dark grey) band. The lower panel of each plot shows the ratio between theoretical predictions and the measurement.}
\label{fig:xsecs}
\end{figure}

\begin{figure}[p!h]
\centering
\includegraphics[width=0.65\textwidth]{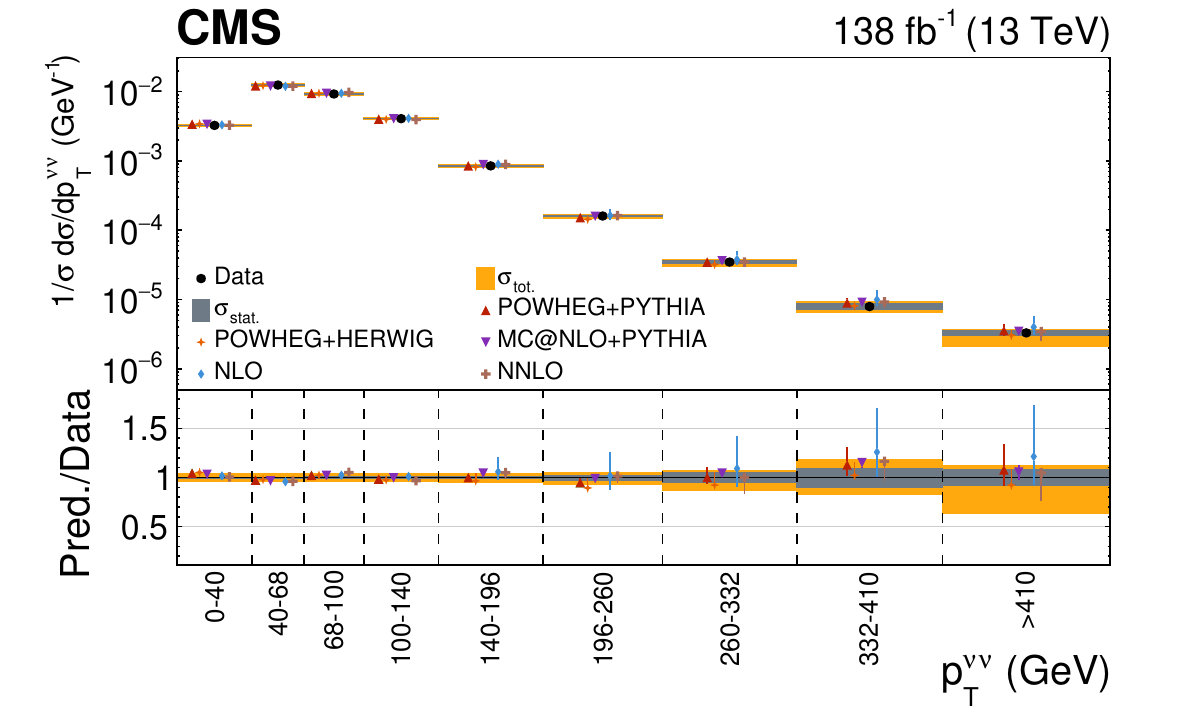} \\
\includegraphics[width=0.65\textwidth]{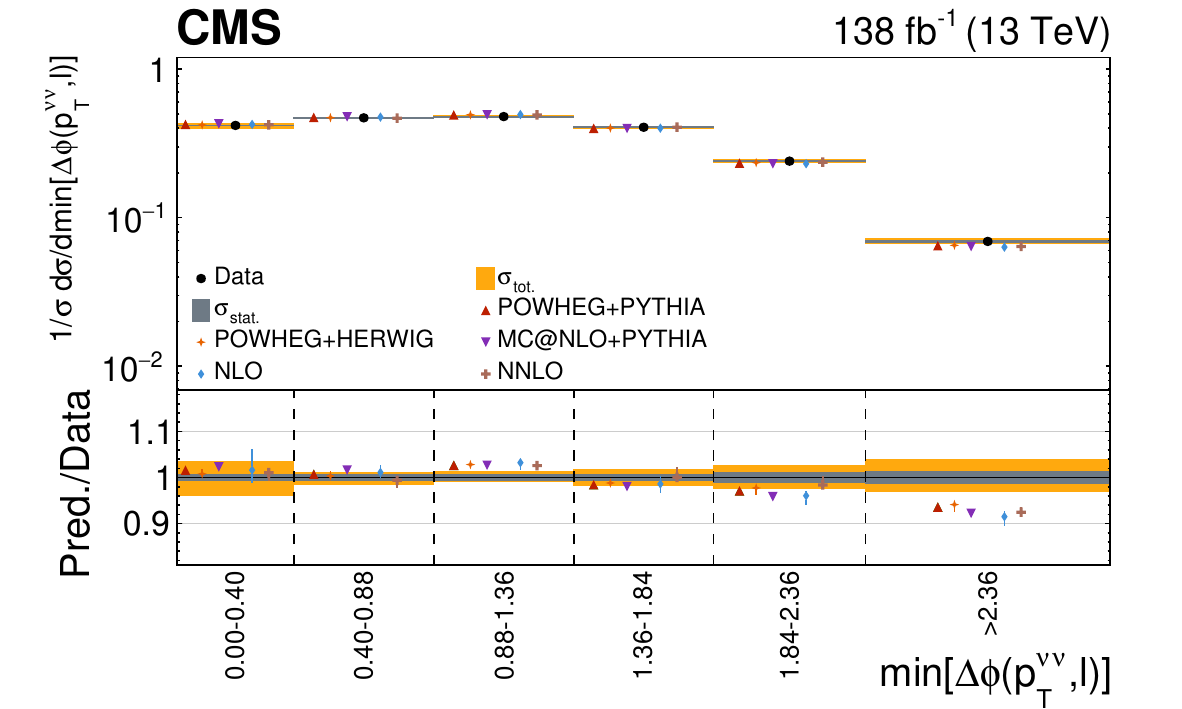} \\
\includegraphics[width=0.65\textwidth]{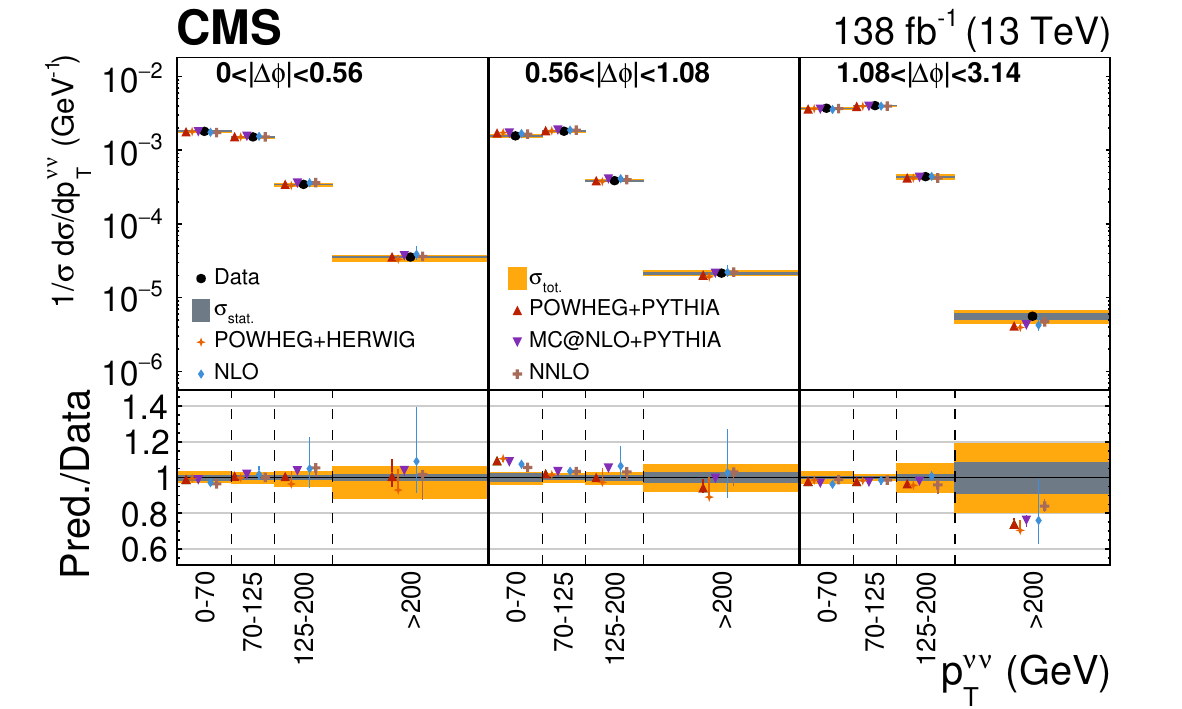}
\caption{The measured normalized differential signal cross sections (black markers) as functions of \pTnunu (upper), \phiNextLepNuNuVec (middle), and double-differential in both observables (lower) are shown. The theoretical predictions from \POWHEG{}+\PYTHIA (dark red), \POWHEG{}+\HERWIG (orange), \MCATNLO{}+\PYTHIA (purple), and the fixed-order NLO (light blue) and NNLO (brown) calculations are compared to the measurement. The cross sections in the last \pTnunu bin of the upper (lower) plot are determined for $\pTnunu>410$ (200)\GeV and normalised to a bin width of 90 (200)\GeV for visualisation purposes. The total (statistical) uncertainty on the measurement is shown as an orange (dark grey) band. The lower panel of each plot shows the ratio between theoretical predictions and the measurement.}
\label{fig:xsecs_norm}
\end{figure}

The differential signal cross sections and their comparisons to the theoretical predictions are shown in Figs.~\ref{fig:xsecs}--\ref{fig:xsecs_norm}. The corresponding \chisq tests for all measurements, both with and without the inclusion of uncertainties on the predictions, are summarized in Table~\ref{tab:chi2tests}, with the corresponding $p$-values~\cite{ParticleDataGroup:2024cfk} given in Table~\ref{tab:chi2tests_pval}.

\begin{table}[ht!]
\centering
\topcaption{Results of the \chisq tests for the absolute and normalized differential cross section measurements for each of the predictions. The \chisq values including the uncertainties on the predictions are given in parentheses.}
\label{tab:chi2tests}
\renewcommand{\arraystretch}{1.15}
\cmsTable{\begin{tabular}{cc@{\hspace{5ex}}ccccc}
    Measurement & \ndf & \multicolumn{5}{c}{\chisq (including uncertainties)} \\
    \cline{3-7}
    & & \POWHEG  & \POWHEG  & \MCATNLO & \multirow{2}{*}{NLO} & \multirow{2}{*}{NNLO} \\[-3pt]
    & & +\PYTHIA & +\HERWIG & +\PYTHIA & & \\ \hline
    \pTnunu & 9 & 6.8 (1.6) & 8.0 (1.0) & 9.0 (3.9) & 7.1 (1.4) & 12.5 (2.5) \\
    \phiNextLepNuNuVec & 6 & 7.3 (1.5) & 7.4 (1.4) & 11.5 (2.7) & 12.3 (2.5) & 11.9 (4.0) \\
    2D & 12 & 18.8 (4.9) & 22.4 (3.7) & 31.6 (8.6) & 24.0 (3.5) & 15.8 (4.2) \\[\cmsTabSkip]
    norm. \pTnunu & 8 & 6.4 (3.8) & 8.3 (3.2) & 6.4 (5.8) & 6.7 (1.9) & 13.3 (5.5) \\
    norm. \phiNextLepNuNuVec & 5 & 7.2 (7.0) & 7.2 (6.7) & 9.2 (8.4) & 11.7 (5.4) & 11.7 (5.8) \\
    norm. 2D  & 11 & 18.1 (16.0) & 21.4 (14.5) & 28.2 (22.8) & 23.1 (7.0) & 14.8 (8.5) \\
\end{tabular}}
\end{table}

\begin{table}[ht!]
\centering
\topcaption{The $p$-value of the \chisq tests for the absolute and normalized cross section measurements for each of the predictions. The resulting $p$-values including uncertainties on the predictions are given in parentheses.}
\label{tab:chi2tests_pval}
\renewcommand{\arraystretch}{1.15}
\begin{tabular}{cccccc}
    Measurement & \multicolumn{5}{c}{$p$-value (including uncertainties)} \\
    \cline{2-6}
    & \POWHEG  & \POWHEG  & \MCATNLO & \multirow{2}{*}{NLO} & \multirow{2}{*}{NNLO} \\[-3pt]
    & +\PYTHIA & +\HERWIG & +\PYTHIA  & & \\ \hline
    \pTnunu & 0.66 (1.00) & 0.53 (1.00) & 0.44 (0.92) & 0.63 (1.00) & 0.19 (0.98) \\
    \phiNextLepNuNuVec & 0.29 (0.96) & 0.28 (0.97) & 0.07 (0.84) & 0.06 (0.87) & 0.07 (0.67) \\
    2D & 0.09 (0.96) & 0.03 (0.99) & $<$0.01 (0.73) & 0.02 (0.99) & 0.20 (0.98) \\[\cmsTabSkip]
    norm. \pTnunu & 0.60 (0.87) & 0.40 (0.92) & 0.60 (0.67) & 0.57 (0.98) & 0.10 (0.70) \\
    norm. \phiNextLepNuNuVec & 0.21 (0.22) & 0.20 (0.24) & 0.10 (0.14) & 0.04 (0.37) & 0.04 (0.33) \\
    norm. 2D & 0.08 (0.14) & 0.03 (0.21) & $<$0.01 (0.02) & 0.02 (0.80) & 0.19 (0.67) \\
\end{tabular}
\end{table}

In general, all predictions are in agreement with the absolute measurements. In the absolute \phiNextLepNuNuVec measurement, small shape differences between the measured cross section and the five predictions are observed. These differences match observations from previous measurements of the azimuthal angle between two leptons $\Delta\phi(\Pell,\Pell)$~\cite{ATLAS:2023gsl,CMS:2018adi, ATLAS:1612.07004, ATLAS:1903.07570, CMS:1601.01107, CMS:1907.03729}, which is strongly correlated with \phiNextLepNuNuVec. Slightly larger \chisqndf values (where \ndf is the number of degrees of freedom) are observed for the 2D differential cross section measurement. While \MCATNLO{}+\PYTHIA overpredicts the cross sections in most of the measured bins for the \pTnunu distribution in comparison with other predictions, its \chisqndf is found to be close to 1 due to the correlated bin uncertainties. Including the uncertainties on the predictions reduces the \chisqndf of most predictions to below unity. Only the \MCATNLO{}+\PYTHIA prediction leads to \chisqndf values significantly above unity for the normalised 2D measurement. The results of the \chisq tests without including uncertainties show that \POWHEG{}+\PYTHIA yields overall the best description for the two one-dimensional absolute differential measurements, and the NNLO calculation shows the best agreement with the 2D absolute result.

For the normalized differential cross sections, shown in Fig.~\ref{fig:xsecs_norm}, the systematic uncertainties are slightly smaller than those for the absolute cross section measurements. This especially holds for the \phiNextLepNuNuVec measurement, where the dominant uncertainty on the absolute differential measurements that arises from lepton reconstruction efficiency is significantly reduced, which emphasizes the shape differences between the measurement and predictions already discussed for the absolute result. Similar to the absolute results, all predictions are in reasonable agreement with the measurements as shown by the obtained \chisq values.

\section{Summary}
\label{sec:summary}

Differential cross section measurements have been presented for the top quark pair production in the dileptonic channel in proton-proton (\pp) collisions at a center-of-mass energy of 13 \TeV using observables based on the dineutrino kinematic properties. The measurements are performed based on the \pp collision data recorded by the CMS detector at the CERN LHC between 2016 and 2018, corresponding to an integrated luminosity of 138\fbinv.

The differential cross sections are measured as functions of the transverse momentum of the dineutrino system \pTnunu, and the minimal azimuthal angle between the dineutrino system and a charged lepton \phiNextLepNuNuVec, as well as both observables in two dimensions. To improve the resolution of the missing transverse momentum, which serves as a measure of \pTnunu in signal events, a dedicated deep neural network regression has been developed. The method significantly improves the resolution of both the magnitude and the azimuthal angle of the missing transverse momentum. The absolute and normalized differential cross section results are obtained based on an unregularized least squares unfolding method.

The differential cross sections are compared to predictions based on Monte Carlo simulation as well as two fixed-order theoretical calculations, corresponding to next-to-leading order and next-to-next-to-leading order (NNLO) accuracy in quantum chromodynamics. A remarkable agreement between the different theoretical predictions and the measured differential cross sections is observed. For both one-dimensional measurements, the best overall description is provided by the two \POWHEG predictions, while for the two-dimensional measurement the best agreement is observed for the NNLO fixed-order calculation. However, the differences between the five predictions are mostly small, such that none of the predictions is significantly disfavored by the measured differential cross sections. These results constitute the first differential cross section measurements based on the dineutrino kinematic properties in top quark pair production.

\begin{acknowledgments}
We congratulate our colleagues in the CERN accelerator departments for the excellent performance of the LHC and thank the technical and administrative staffs at CERN and at other CMS institutes for their contributions to the success of the CMS effort. In addition, we gratefully acknowledge the computing centers and personnel of the Worldwide LHC Computing Grid and other centers for delivering so effectively the computing infrastructure essential to our analyses. Finally, we acknowledge the enduring support for the construction and operation of the LHC, the CMS detector, and the supporting computing infrastructure provided by the following funding agencies: SC (Armenia), BMBWF and FWF (Austria); FNRS and FWO (Belgium); CNPq, CAPES, FAPERJ, FAPERGS, and FAPESP (Brazil); MES and BNSF (Bulgaria); CERN; CAS, MoST, and NSFC (China); MINCIENCIAS (Colombia); MSES and CSF (Croatia); RIF (Cyprus); SENESCYT (Ecuador); ERC PRG, TARISTU24-TK10 and MoER TK202 (Estonia); Academy of Finland, MEC, and HIP (Finland); CEA and CNRS/IN2P3 (France); SRNSF (Georgia); BMFTR, DFG, and HGF (Germany); GSRI (Greece); NKFIH (Hungary); DAE and DST (India); IPM (Iran); SFI (Ireland); INFN (Italy); MSIT and NRF (Republic of Korea); MES (Latvia); LMTLT (Lithuania); MOE and UM (Malaysia); BUAP, CINVESTAV, CONACYT, LNS, SEP, and UASLP-FAI (Mexico); MOS (Montenegro); MBIE (New Zealand); PAEC (Pakistan); MES, NSC, and NAWA (Poland); FCT (Portugal); MESTD (Serbia); MICIU/AEI and PCTI (Spain); MOSTR (Sri Lanka); Swiss Funding Agencies (Switzerland); MST (Taipei); MHESI (Thailand); TUBITAK and TENMAK (T\"{u}rkiye); NASU (Ukraine); STFC (United Kingdom); DOE and NSF (USA).

\hyphenation{Rachada-pisek} Individuals have received support from the Marie-Curie program and the European Research Council and Horizon 2020 Grant, contract Nos.\ 675440, 724704, 752730, 758316, 765710, 824093, 101115353, 101002207, 101001205, and COST Action CA16108 (European Union); the Leventis Foundation; the Alfred P.\ Sloan Foundation; the Alexander von Humboldt Foundation; the Science Committee, project no. 22rl-037 (Armenia); the Fonds pour la Formation \`a la Recherche dans l'Industrie et dans l'Agriculture (FRIA) and Fonds voor Wetenschappelijk Onderzoek contract No. 1228724N (Belgium); the Beijing Municipal Science \& Technology Commission, No. Z191100007219010, the Fundamental Research Funds for the Central Universities, the Ministry of Science and Technology of China under Grant No. 2023YFA1605804, and the Natural Science Foundation of China under Grant No. 12061141002 (China); the Ministry of Education, Youth and Sports (MEYS) of the Czech Republic; the Shota Rustaveli National Science Foundation, grant FR-22-985 (Georgia); the Deutsche Forschungsgemeinschaft (DFG), among others, under Germany's Excellence Strategy -- EXC 2121 ``Quantum Universe" -- 390833306, and under project number 400140256 - GRK2497; the Hellenic Foundation for Research and Innovation (HFRI), Project Number 2288 (Greece); the Hungarian Academy of Sciences, the New National Excellence Program - \'UNKP, the NKFIH research grants K 131991, K 133046, K 138136, K 143460, K 143477, K 146913, K 146914, K 147048, 2020-2.2.1-ED-2021-00181, TKP2021-NKTA-64, and 2021-4.1.2-NEMZ\_KI-2024-00036 (Hungary); the Council of Science and Industrial Research, India; ICSC -- National Research Center for High Performance Computing, Big Data and Quantum Computing, FAIR -- Future Artificial Intelligence Research, and CUP I53D23001070006 (Mission 4 Component 1), funded by the NextGenerationEU program (Italy); the Latvian Council of Science; the Ministry of Education and Science, project no. 2022/WK/14, and the National Science Center, contracts Opus 2021/41/B/ST2/01369, 2021/43/B/ST2/01552, 2023/49/B/ST2/03273, and the NAWA contract BPN/PPO/2021/1/00011 (Poland); the Funda\c{c}\~ao para a Ci\^encia e a Tecnologia, grant CEECIND/01334/2018 (Portugal); the National Priorities Research Program by Qatar National Research Fund; MICIU/AEI/10.13039/501100011033, ERDF/EU, "European Union NextGenerationEU/PRTR", and Programa Severo Ochoa del Principado de Asturias (Spain); the Chulalongkorn Academic into Its 2nd Century Project Advancement Project, the National Science, Research and Innovation Fund program IND\_FF\_68\_369\_2300\_097, and the Program Management Unit for Human Resources \& Institutional Development, Research and Innovation, grant B39G680009 (Thailand); the Kavli Foundation; the Nvidia Corporation; the SuperMicro Corporation; the Welch Foundation, contract C-1845; and the Weston Havens Foundation (USA).
\end{acknowledgments}

\bibliography{auto_generated}
\cleardoublepage \appendix\section{The CMS Collaboration \label{app:collab}}\begin{sloppypar}\hyphenpenalty=5000\widowpenalty=500\clubpenalty=5000\input{TOP-24-001-public-authorlist.tex}\end{sloppypar}
\end{document}

%% file: TOP-24-001-public-authorlist.tex
\cmsinstitute{Yerevan Physics Institute, Yerevan, Armenia}
{\tolerance=6000
V.~Chekhovsky, A.~Hayrapetyan, V.~Makarenko\cmsorcid{0000-0002-8406-8605}, A.~Tumasyan\cmsAuthorMark{1}\cmsorcid{0009-0000-0684-6742}
\par}
\cmsinstitute{Institut f\"{u}r Hochenergiephysik, Vienna, Austria}
{\tolerance=6000
W.~Adam\cmsorcid{0000-0001-9099-4341}, J.W.~Andrejkovic, L.~Benato\cmsorcid{0000-0001-5135-7489}, T.~Bergauer\cmsorcid{0000-0002-5786-0293}, K.~Damanakis\cmsorcid{0000-0001-5389-2872}, M.~Dragicevic\cmsorcid{0000-0003-1967-6783}, P.S.~Hussain\cmsorcid{0000-0002-4825-5278}, M.~Jeitler\cmsAuthorMark{2}\cmsorcid{0000-0002-5141-9560}, N.~Krammer\cmsorcid{0000-0002-0548-0985}, A.~Li\cmsorcid{0000-0002-4547-116X}, D.~Liko\cmsorcid{0000-0002-3380-473X}, I.~Mikulec\cmsorcid{0000-0003-0385-2746}, J.~Schieck\cmsAuthorMark{2}\cmsorcid{0000-0002-1058-8093}, R.~Sch\"{o}fbeck\cmsAuthorMark{2}\cmsorcid{0000-0002-2332-8784}, D.~Schwarz\cmsorcid{0000-0002-3821-7331}, M.~Sonawane\cmsorcid{0000-0003-0510-7010}, W.~Waltenberger\cmsorcid{0000-0002-6215-7228}, C.-E.~Wulz\cmsAuthorMark{2}\cmsorcid{0000-0001-9226-5812}
% ...
\par}
\vskip\cmsinstskip
\dag:~Deceased\\
$^{1}$Also at Yerevan State University, Yerevan, Armenia\\
$^{2}$Also at TU Wien, Vienna, Austria\\
% ...
